\documentclass[aps,prresearch,twocolumn,superscriptaddress,bibnotes, longbibliography, floatfix]{revtex4-2}%
\usepackage{graphicx}
\usepackage{bm,color}
\usepackage{amsmath,amssymb,dsfont,amstext,amsfonts}
\usepackage{extarrows}
\usepackage[colorlinks=true,linkcolor=blue,urlcolor=blue,citecolor=blue]{hyperref}
\usepackage{xcolor}
\usepackage{wasysym}
\usepackage{mathtools}
\usepackage{bbold}
\usepackage{import}
\usepackage{subfigure}

\usepackage[normalem]{ulem} 
\usepackage{color}

\def\bra#1{\mathinner{\langle{#1}|}}
\def\ket#1{\mathinner{|{#1}\rangle}}

\begin{document}
\title{Localization and topological signatures under periodic twisting}

\author{James Walkling}
\email{jamwalk@pks.mpg.de}  
\affiliation{Max-Planck-Institut für Physik komplexer Systeme, Nöthnitzer Straße 38, 01187 Dresden, Germany}
\affiliation{TCM Group, Cavendish Laboratory, University of Cambridge, JJ Thomson Avenue, Cambridge CB3 0HE, United Kingdom\looseness=-1}

\author{Antonio \v{S}trkalj}
\email{astrkalj@phy.hr}
\thanks{These authors contributed equally to this work.}
\affiliation{Department of Physics, Faculty of Science, University of Zagreb, Bijeni\v{c}ka c. 32, 10000 Zagreb, Croatia}
\affiliation{TCM Group, Cavendish Laboratory, University of Cambridge, JJ Thomson Avenue, Cambridge CB3 0HE, United Kingdom\looseness=-1}

\author{F.~Nur \"Unal}
\email{f.unal@bham.ac.uk}
\thanks{These authors contributed equally to this work.}
\affiliation{School of Physics and Astronomy, University of Birmingham, Edgbaston, Birmingham B15 2TT, United Kingdom\looseness=-1}
\affiliation{TCM Group, Cavendish Laboratory, University of Cambridge, JJ Thomson Avenue, Cambridge CB3 0HE, United Kingdom\looseness=-1}

\begin{abstract}

We theoretically explore a dynamical generalization of the Aubry-André model in two dimensions formed by superimposing two square-lattice potentials. Motivated by the rich physics emerging at different twist angles between the two lattices at equilibrium, we introduce periodic twisting by continuously rotating one of the lattices with respect to the other in the plane. We demonstrate that the distinct time-dependent twisting in this system gives rise to an intricate form of periodic multi-frequency driving that changes with the distance from the rotation axis. We find that the incommensurate nature of the potential no longer plays the pivotal role as it does in the static case. Rather, the tunnelling can be understood in terms of a local, spatially varying dynamical localization effect, which we show to yield ring-shaped states localized within the bulk that have interesting transport signatures. Quantifying the eigenstates with the Bott index and local Chern marker, we find that there is a zoo of states with non-trivial topological signatures, the most ubiquitous of which result in relatively uniform ring-shaped regions of the Chern marker. We investigate the origin of these effects from various angles and identify that hybridization between different delocalized ring states plays a vital role. Lastly, we discuss possible experimental realizations in quantum simulation settings. Our results open a new avenue of investigation with periodic twisting inducing a spatially varying multi-frequency drive.
\end{abstract}

\maketitle

\section{Introduction}
Multi-layer systems have recently emerged as a fertile playground for studying exotic physical phenomena. On one hand, it has instigated the rising field of twistronics, where two-dimensional lattices are stacked with relative lattice mismatches, with different commensurate or incommensurate twist angles that can form moir\'e supercells~\cite{BM2011}. This induces novel properties such as flat-band superconductivity~\cite{Cao2018, Lu2019, Stepanov2020}, Mott insulating behaviour~\cite{Song2023, Deng2025,Cao2018ins}, non-trivial topologies~\cite{Rosa2021,Song2022} or ferroelectricity~\cite{Yasuda21_Science,woods2021_NatComm,Klein2023} to name a few. 
On the other hand, superimposing two lattices incommensurately can produce aperiodic structures~\cite{Rosa2021,ViebahnSchneider19_PRL_QC,KozlovLevitov2023_arxiv}, which are deterministic but lack translational invariance and therefore lie between uniform and fully disordered systems. Such quasiperiodic systems offer intriguing localization properties such as metal-to-insulator transition in one and two dimensions~\cite{AA1980,Jitomirskaya1999,Szabo2020,Sbroscia2020}, stable many-body localized phases~\cite{Strkalj2022}, critical eigenmodes and fractal energy spectrum ~\cite{Kohmoto1983, Ostlund1983, Kohmoto1984,Jagannathan2021}, as well as new kinds of topological phenomena with potentially no crystalline counterpart~\cite{Bandres16_PRX,Johnstone2022}. 

In this regard, quantum simulators provide unique opportunities to realize quasiperiodic models and investigate new regimes in experiments~\cite{Bordia2017,ViebahnSchneider19_PRL_QC,Bandres16_PRX,Strkalj2020}. 
One stronghold of these platforms, such as ultracold atoms, is that they unlock a plethora of dynamical considerations allowing for exploring non-equilibrium phenomena involving periodic driving, sudden quenches, or adiabatic modulations of parameters~\cite{Cooper2019}. 
In the case of quasiperiodic systems, such adiabatic modulations give rise to exotic localization and topological properties, where the latter can often be connected to higher-dimensional topologies such as the Thouless pump in one dimension~\cite{Marra20_PRB_qperThouless,KrausZilberberg12_PRL_1Dqc, Rosa2021,Gottlob25_PRXq,BaiWeld2025} or the four-dimensional quantum Hall effect in two-dimensional quasicrystals~\cite{KrausZilberberg13_PRL_2Dqc}. 

In this paper, we introduce a distinctive out-of-equilibrium scenario, namely the time-dependent periodic twisting in a two-dimensional (2D) quasiperiodic system formed by superimposing two lattice potentials~\cite{Jagannathan2013,
ViebahnSchneider19_PRL_QC, JiaZheng2022_RBG, HuangLiu2019_PRB,Sbroscia2020,Szabo2020,Strkalj2022,Gottlob2023}.
For fixed twist angles (the static case) considered previously, the incommensurability between two potentials eventually yields a localization transition of most of the states in the spectrum for strong enough potentials~\cite{Strkalj2022,Szabo2020}. However, periodically modulated systems are known to feature behaviour distinct from equilibrium counterparts~\cite{Rudner20_NatPhysRev,Kitagawa10_PRB,Rudner13_PRX,Titum2016_PRX_AFAI,Nathan2019_PRB_AFI,slager2024_NatCom_ADS}. We here demonstrate how the multi-frequency nature of the periodic twisting gives rise to unique localization and topological effects emerging on different length scales.

We start with a 2D generalization of the paradigmatic Aubry-André (AA) model~\cite{AA1980}.
This can be realized, for example, by overlapping two square optical lattices with incommensurate wavelengths~\cite{ViebahnSchneider19_PRL_QC,HuangLiu2019_PRB,Sbroscia2020,Szabo2020,Strkalj2022}. We consider the first (primary) square lattice to have a stronger potential admitting a tight-binding description, where the second one is weaker and induces on-site potentials. Quasiperiodic nature of the AA model together with different static twist angles between the two layers has been previously shown to result in intriguing properties, for non-interacting as well as interacting particles, where the localization phenomena have been at the focus of research~\cite{HuangLiu2019_PRB,Sbroscia2020,Szabo2020,Strkalj2022}. 
Upon introducing periodic twisting of the perturbing lattice in this manuscript, we investigate the localization transition, or lack thereof. We first demonstrate that the time-dependent twisting in our system induces a spatially-changing periodic modulation effect. This introduces a novel type of driving, an intrinsically multi-frequency drive that is governed by varying harmonics at different length scales. While on one hand the incommensurate spatial character of the twisting promotes an effective quasiperiodic potential, we find that, on the other hand, the time-periodic nature of the drive dominates this process and gives rise to an inhomogeneous dynamical localization effect.

\section{Model}
%
\begin{figure}[t]
    \centering
    \includegraphics[width=1\linewidth]{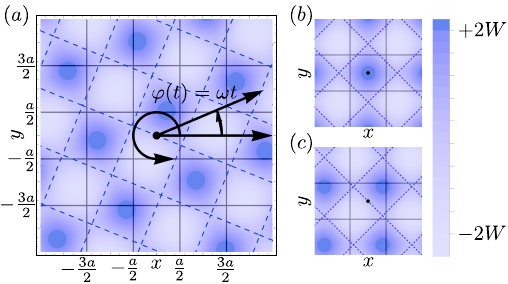}
    \caption{
    (a) A contour plot of the on-site potential with strength $W$ at time $t$, with dashed blue lines marking the zero potential lines. The solid lines depict the tight binding lattice with sites on the vertices and lattice constant $a=1$. The perturbing lattice is being rotated counter-clockwise (black arrows) with an angle $\varphi(t)$ about an axis perpendicular to the lattice plane.
    (b) and (c) show the on-site potential at $t=0$ for two different offsets $\Theta=0$ and $\Theta=\pi/2$, respectively.}
    \label{fig:potentialexample}
\end{figure}
The Hamiltonian of our system is given by
\begin{equation}
    \hat{H} (t)= -J \sum_{\langle \mathbf{n,n'}\rangle} c^\dagger_{\mathbf{n}}c_{\mathbf{n}'} + \sum_{\mathbf{n}} V_{\mathbf{n}}(t) c^\dagger_\mathbf{n} c_\mathbf{n},
    \label{eqn:Hamiltonian}
\end{equation}
where $J$ is the tunnelling amplitude between nearest-neighbour pairs of sites labelled by ${\langle \mathbf{n},\mathbf{n}'\rangle}$ in the primary lattice and $c_{\mathbf{n}}^{(\dagger)}$ is the annihilation (creation) operator. Throughout the text, we express all energies in units of $J$ and set the tight-binding lattice constant $a=1$ together with $\hbar=c=1$. The time-dependent on-site term $V_{\mathbf{n}}(t)$ has the form
\begin{equation}
    V_\mathbf{n}(t) = W \big[ \cos(2\pi \beta u_\mathbf{n}(t) - \Theta)+\cos(2\pi\beta v_\mathbf{n}(t) - \Theta) \big] \, ,
    \label{eqn:potential}
\end{equation}
created by considering a secondary square lattice potential with strength $W$, and $\Theta$ is the phase that spatially translates the potential in the plane with respect to the underlying lattice. We denote the ratio of the two lattice constants with $\beta$, which determines the spatial frequency in Eq.~\eqref{eqn:potential} of the on-site potential---not to be confused with the periodic driving frequency later. 
The variables $(u_\mathbf{n},v_\mathbf{n})$, are related to a lattice site $\mathbf{n}=(x_\mathbf{n},y_\mathbf{n}) \in \mathbb{Z}^2$ by a rotation matrix $\mathbf{R}_{\varphi=\omega t}$ such that 
\begin{align}
\begin{pmatrix}
    u_\mathbf{n}(t) \\
    v_\mathbf{n}(t) 
\end{pmatrix} = \begin{pmatrix}
\cos(\omega t) & \sin(\omega t) \\
-\sin(\omega t) & \cos(\omega t) 
\end{pmatrix} \begin{pmatrix}
    x_\mathbf{n} \\
    y_\mathbf{n} 
\end{pmatrix} \, ,
\end{align}
in which the rotation frequency of the potential $V_\mathbf{n}(t)$ is given by $\omega$. For $\omega >0$, the perturbing lattice rotates counter-clockwise. 

We demonstrate the spatial profile of the potential in Fig.~\ref{fig:potentialexample}(a). Two different choices of the phase $\Theta$ are given in Fig.~\ref{fig:potentialexample}(b) and (c), showing how it changes the centre of rotation and the symmetry of the potential. 
The $\Theta=0$ case has inversion symmetry with respect to the  rotation axis, while $\Theta=\pi/2$ breaks it. As we discuss in the following sections, this has a profound impact on the localization and topological properties of the system.
In the rest of the paper, we concentrate on these two values of $\Theta$.

In order to solve the time-dependent problem, we first split the contributions to the on-site potential as $V_\mathbf{n}(t) = \langle V_\mathbf{n} \rangle + W \sum_{\nu \neq 0} s_{\nu,\mathbf{n}} \exp(i \nu \omega t)$, where the first term is averaged over a single period of rotation and, therefore, time independent, while the second term is time dependent and written in terms of the finite-frequency Fourier components.
Following this decomposition, it is convenient to shift the time dependence from on-site terms to the hopping terms in Eq.~\eqref{eqn:Hamiltonian}~\cite{Eckardt2017}. We use a gauge transformation of the form $U(t)=\exp\left(-i\sum_\mathbf{n} \int^t_{t_0} \text{d} t' V_\mathbf{n}(t') c^\dagger_\mathbf{n} c_\mathbf{n} \right)$, where  we take $t_0=0$ without loss of generality (see Appendix~\ref{section:gauge} for details). The Hamiltonian in this lattice frame is found as
\begin{align}
    \hat{H}' (t) = - \sum_{\langle \mathbf{n},  \mathbf{n}' \rangle}  J_{\mathbf{n} \mathbf{n}'}(t) c^\dagger_\mathbf{n}c_{\mathbf{n}'} +\sum_\mathbf{n} \langle V_\mathbf{n} \rangle c^\dagger_\mathbf{n} c_\mathbf{n}  \,,
    \label{eqn:transformedH}
\end{align}
where the local time-dependent tunnelling amplitudes can be expressed in terms of the Fourier components $s_{\nu,\mathbf{n}}$ of the on-site potential as
\begin{align}
    J_{\mathbf{n} \mathbf{n}'}(t)= J\exp\bigg\{\frac{-i W}{\omega} \,\sum_{\nu=-\infty}^{\infty} \frac{ s_{\nu,\mathbf{n}}-s_{\nu,\mathbf{n}'}}{i\nu} \exp(i\nu \omega t) \bigg\} \, .
\label{eqn:Jefft}
\end{align}
We tabulate their explicit form in Appendix~\ref{section:Spectrum} for the two different values of $\Theta$ mentioned above. 

The stroboscopic time evolution of the system is calculated as $U(T,0) = \mathfrak{T}\exp\{-i\int_0^T \hat{H}'(t)dt\}$ with time ordering $\mathfrak{T}$ and period $T=2\pi/\omega$. The Floquet Hamiltonian $H_F$ reproducing this stroboscopic evolution is given by
\begin{equation}\label{eq:U}
    U(T,0) = \exp(-i H_F T),
\end{equation}
where the $m$th quasienergy $\varepsilon^m$ and Floquet states $\psi^m$ are defined via $H_F\psi^m=\varepsilon^m\psi^m$. 
We investigate the evolution of the system by analytically studying the asymptotic behaviour of the perturbing potential and stroboscopic dynamics, as well as by numerically solving for the Floquet Hamiltonian using Runge-Kutta methods. We find that the properties of the quasienergy eigenstates of $H_F$ is dramatically different depending on the symmetry of the time-dependent potential given by $\Theta$ in Eq.~\eqref{eqn:potential}. In the subsequent sections, we analyse the localization properties and spectrum for the $\Theta=\pi/2$ case since it displays a richer range of behaviour, while we relegate the discussion of the case with $\Theta=0$ to Appendix~\ref{sec:evensymloc}.

\section{Nature of the Driving}
\begin{figure}[t!]
    \centering    \includegraphics[width=1\linewidth]{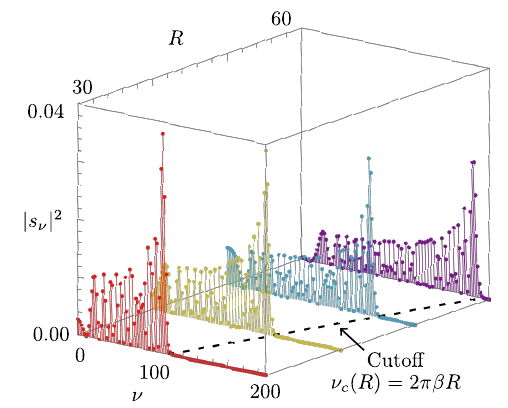}
    \caption{A plot of the Fourier coefficients, $s_{\nu,\mathbf{n}}$ of the on-site potential in Eq.~\eqref{eqn:potential} with $\Theta=\pi/2$ for $R=|\mathbf{n}|=30,40,50,60$ sites away from the origin (rotation axis). Periodic twisting induces a local multi-frequency drive with a spectral peak at larger frequencies further from the axis of rotation. The values at which $s_\nu$ drop to zero follow a linear trend in $R$ (dashed line) in agreement with our definition of $\nu_{\rm c}$ in the main text. }
    \label{fig:drivingex}
\end{figure}

It is known that the out-of-equilibrium behaviour in periodically driven systems can be quite different in the low- and high-frequency regimes which can be captured by certain effective descriptions~\cite{Eckardt2017,Vogl2020,Rodriguez-Vega2018}. The two regimes are 
usually defined with respect to the energy scales of the undriven system, where in our case the bandwidth of the single-particle spectrum of the underlying square lattice sets an energy scale $\Omega_c=8J$. 
However, the continuous twisting described in this work induces a unique periodic drive that involves an extensive number of harmonics which can reach into both regimes.
While the drive originates from the rotation of an on-site potential landscape around a single axis with constant rotation frequency $\omega$, different lattice sites feel different linear velocities. This excites a whole spectrum of local driving frequencies (harmonics) in the system in a way that is spatially inhomogeneous. We find that the spectrum at a given site has a maximum harmonic $\nu_c(R)$ which depends on the distance $R$ from the axis of rotation. 
We demonstrate this in Fig.~\ref{fig:drivingex} by using the Fourier decomposition of the local driving $V_\mathbf{n}(t)$ from Eq.~\eqref{eqn:potential}, see Appendix~\ref{section:Spectrum} for further details where we explicitly calculate the Fourier coefficients.

There are two main features that follow from the analysis of Fig.~\ref{fig:drivingex}. 
First, we notice that the frequency spectrum gets broader for sites further away from the axis of rotation, i.e.~with a maximum harmonic, $\nu_c$, that increases linearly with distance $R$. This relationship can be understood heuristically: Along a circumference at a given $R$, the number of local minima and maxima in the potential scales like $R$. The $\nu_c$ locally at a given site will scale linearly with the number of local minima and maxima that pass through it in one rotation of the on-site potential landscape, hence it scales linearly with $R$. Consequently, even for low rotation frequencies $\omega < \Omega_{\rm c}$, we observe that parts of the lattice further away from the axis of rotation are driven with local frequencies that are in the high-frequency regime.
Second, we find that the squared amplitude of Fourier coefficients can, to a coarse approximation, be described as uniform before a cut-off at $\nu_c$ as we will demonstrate below. Crucially, this decreases for far-away sites as $|s_{\nu}|^2 \propto R^{-1}$ (see Appendix~\ref{section:Spectrum}). These two features of the local driving in both its frequency and amplitude are ultimately what gives rise to the characteristic features of our model and distinguishes it from extensive prior work on periodically driven systems. 

Following the behaviour of $|s_{\nu}|^2$ presented in Fig.~\ref{fig:drivingex} and the local multi-frequency character discussed above, we now turn to the analysis of the time-dependent nearest-neighbour tunnelling amplitudes in Eq.~\eqref{eqn:Jefft}.
For convenience, we express it as $J_{\mathbf{n} \mathbf{n}'}(t)=J \exp\{-i\phi_{\mathbf{n} \mathbf{n}'}(t)\}$. By understanding the behaviour of the phase $\phi_{\mathbf{n} \mathbf{n}'}(t)$ with distance, we can quantify the effect of the driving on tunnelling between different sites.
Defining $\gamma_\mathbf{n} = 2\pi \beta R_\mathbf{n}$ for a site $\mathbf{n}$ with polar coordinates $(R_\mathbf{n}, \theta_\mathbf{n})$, as we show in Appendix~\ref{sec:spatvarhop}, the time-varying Peierls phase can be expressed via Fourier modes as
\begin{equation}
    \phi_{\mathbf{n} \mathbf{n}'}(t) = \frac{W}{\omega}\sum_{\nu =\text{odd}} \alpha_{\nu,\mathbf{n} \mathbf{n}'} \sin(\nu \omega t + \Delta_{\nu, \mathbf{n} \mathbf{n}'}) \, ,
    \label{eqn:phasetime}
\end{equation}
with
\begin{equation}
    \alpha_{\nu,\mathbf{n} \mathbf{n}'} = \frac{2\sqrt{2}}{\nu} \Big|\mathcal{J}_\nu(\gamma_\mathbf{n})+\mathcal{J}_\nu(\gamma_{\mathbf{n}'}) \exp(i\nu(\theta_{\mathbf{n}}-\theta_{\mathbf{n}'})) \Big| \, ,
    \label{eqn:alphas}
\end{equation}
and $\Delta_{\nu, \mathbf{n} \mathbf{n}'}$ being a site-dependent phase that is irrelevant for the physics we discuss in the following.
$\mathcal{J}_\nu(\gamma_\mathbf{n})$ denote the Bessel functions of the first kind. 
Note that only the odd modes are non-zero for the Fourier coefficients of driving with $\Theta=\pi/2$ (see Appendix~\ref{sec:spatvarhop} for details). As anticipated, the strength of the renormalized hopping amplitudes is controlled by a dimensionless driving parameter $W/\omega$ in Eq.~\eqref{eqn:Jefft}~\cite{Eckardt2017}.

Furthermore, due to the asymptotic behaviour of the Bessel functions $\mathcal{J}_\nu$ in Eq.~\eqref{eqn:alphas}, we find that far from the axis of rotation, the root mean square (RMS) of the phase decays with distance $R$ as
\begin{equation}    \label{eq:RMS}
    \sqrt{\langle \phi_{\mathbf{n} \mathbf{n}'}^2\rangle} \sim \frac{W}{\omega} \frac{1}{\sqrt{\beta R}} \, .
\end{equation} 
For $R \gg \beta^{-1} W^2 \omega^{-2} $, the RMS phase no longer winds full revolutions, and so the driving has a limited effect on the bare tunnelling amplitudes far away from the axis of rotation. 
Thus, we expect the strongest effects of the driving to occur near the axis of rotation, while far away from it, the hopping amplitude modulation induced by the external drive becomes much less pronounced.  

Note also that although the presence of a broad spectrum of the drive at sites further away from the axis of rotation (c.f.~Fig.~\ref{fig:drivingex}) makes our driving much more complicated, we find that remarkably we can often describe the key behaviour by considering a single effective driving frequency albeit a spatially-dependent one. 
The main reason behind such approximation is the fact that the amplitude $\alpha_{\nu,\mathbf{n} \mathbf{n}'}$ in Eq.~\eqref{eqn:alphas} decays algebraically with $\nu$ and therefore higher harmonics have a lesser impact on the phase $\phi_{\mathbf{n} \mathbf{n}'}$ in Eq.~\eqref{eqn:phasetime}.
\begin{figure}[t!]
  \centering
  \includegraphics[width=1\linewidth]{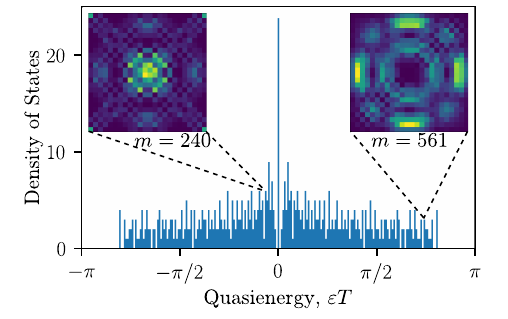}
  \caption{Density of states for a 24 $\times$ 24 lattice with $\Theta=\pi/2$, $W/J=8$ and $\omega/J=9$, as a function of dimensionless quasienergy $\varepsilon T$. The histogram bins are of width $\delta \varepsilon T \approx 0.025$. 
  The insets are spatial probability distribution of two states with ordinal numbers $m=240$ and $m=561$. The former occupies the center of the system, while the latter has most of its support lying on the ring-shaped outer regions.}
  \label{fig:DOSplots2}
\end{figure}

{\it Density of states}--- The subtle spatially-dependent, multi-frequency nature of the driving can play an important role inducing rich localization or topological properties as phases and amplitudes are tuned~\cite{Viebahn2ToneDrive2022,PRajagopalWeld19_PRL_phasonic,Wang2frequency2023}. 
The first signature of these effects emerges in the density of states, which undergoes notable modifications as the driving parameters are modified. 
We show the numerically obtained density of states for a driven system with $\Theta=\pi/2$ in Fig.~\ref{fig:DOSplots2} as a function of dimensionless quasienergy $\varepsilon T$. In contrast to what one would expect in the case of a static cosine potential, the density of states is higher around zero quasienergy.
Further, a prominent peak at $\varepsilon T = 0$ appears. 
Those states are the result of the reflection anti-symmetry of the chosen potential, i.e.~the fact that the potential reverses sign under a $\pi$ rotation as can be seen in Fig.~\ref{fig:potentialexample}, and are present also in the static case. 
It can be shown that the number of zero-energy modes scales with the length of the system's diagonal in the static case for particular twist angles, see Appendix.~\ref{sec:Staticsol} where we discuss the vanishing and infinite potential limits explicitly. 
This degeneracy, surprisingly, survives also in the driven case, and we observe that it remains robust to any finite values of $W$ and $\omega$~\cite{KohmotoSutherland86_PRB,Oktel21_PRB}.

We show the spatial density profiles of two states chosen in the spectrum in the insets of Fig.~\ref{fig:DOSplots2}. The first state is located close to the middle of the quasienergy spectrum, at $\varepsilon T \approx -0.2$ and the second is chosen from the skirts of the spectrum and has $\varepsilon T \approx 2.33$. These states have contrasting spatial profiles, namely, the spatial density of the former is located closer to the axis of rotation, while the latter is depleted there and is supported more towards the edges of the sample. We find that this difference induces interesting localization properties as will be explored next.

\section{localization}
Motivated by the discussion above, we turn to a detailed study of localization properties of the system. To probe localization, we employ the inverse participation ratio (IPR)~\cite{Edwards1972}, defined for each (quasi)energy eigenstate as
\begin{equation}
    \text{IPR}^m = \sum_{\mathbf{n}} |\psi^m_\mathbf{n}|^4, 
    \label{eqn:IPR}
\end{equation}
where the sum goes over all sites and $\psi^m_\mathbf{n}$ is the component of the normalized $m$-th eigenstate in the discrete position basis $\mathbf{r}_\mathbf{n}$. 
The IPR takes a maximum value of 1 for a wavefunction localized on a single site. For a wavefunction homogeneously extended over a whole $N \times N$ lattice, the IPR goes to zero in the thermodynamic limit as ${\rm IPR} \sim 1/N^2$. Hence, in our finite systems, we expect the delocalized states to have low values of the IPR. 
Note that the IPR gives us a measure of localization in both static and driven systems, which makes it an ideal tool for comparing the properties in both cases.
Furthermore, it is important to note that the IPR is ill-defined for degenerate states, since one has the freedom to take less or more localized eigenbases within the degenerate manifold.

\subsection{Static Potential}
Before investigating the effects of periodic twisting, we briefly summarize the localization properties of the static case to draw a clearer contrast, which are investigated to a large extent e.g.~in Refs.~\cite{Szabo2020,Strkalj2022,Duncan2024,HuangLiu2019_PRB} focusing on various aspects.
When $\omega=0$ and the twist angle $\varphi=0$ (cf.~Fig.~\ref{fig:potentialexample}(a)), there is a sharp localization transition captured by high IPR values at the self-duality point $W/J=2$ for any value of the phase shift $\Theta$ if $\beta$ is irrational. %
This follows directly from the fact that the particle's motion in $x$- and $y$-direction is separable, and therefore the localization properties are inherited from the 1D Aubry-Andr\'{e} model~\cite{AA1980}. 
On the other hand, for finite twist angles, the localization properties change. For example, at $\Theta=\pi/4$, the spectrum no longer hosts a localization transition of all eigenstates at the same $W/J=2$, revoking the self-duality point, and the subdimensional extended states persist to arbitrary large $W$~\cite{Szabo2020,Strkalj2022}.

\begin{figure}
    \centering
    \includegraphics[width=1\linewidth]{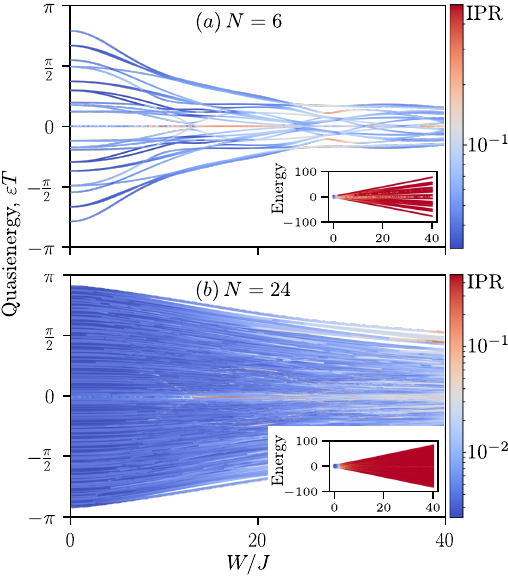}
    \caption{The dimensionless quasienergy, $\varepsilon T$, as a function of potential strength $W/J$ for $\Theta=\pi/2$ and $\omega/J=9$ for $N\times N$ lattices with (a) $N=6$ and (b) $N=24$. Each eigenstate is coloured by its IPR, defined in Eq.~\eqref{eqn:IPR}. The insets in each panel show the corresponding energy spectra for $\omega=0$ over the same range of $W/J$ demonstrating the delocalizing effect of the drive.}
    \label{fig:sinspecexamples}
\end{figure}

\subsection{Driven Potential with \texorpdfstring{$\Theta=\pi/2$}{Theta=pi/2}}
\label{sec:oddsymloc}
Let us now turn on the external driving in the form of periodic twisting shown in Eq.~\eqref{eqn:potential}. Throughout the paper, if not specified otherwise, we take $\beta$ to be equal to the inverse of the golden mean, i.e.~$\beta =\beta_0= 2/(1+\sqrt{5})$, and concentrate on the $\Theta=\pi/2$ case (cf.~Fig~\ref{fig:potentialexample}(c)), while the $\Theta=0$ case is discussed in Appendix~\ref{sec:evensymloc}. The former has no non-trivial rotational symmetries which would bring the potential in its original position before a full $2 \pi$ revolution, so the Floquet frequency is given by the rotational frequency $\Omega=\omega$. 
As already mentioned, the potential reverses sign under a $\pi$ rotation and, consequently, its average over a single period vanishes, $\langle V_\mathbf{n} \rangle =0$. 
We present the numerically calculated quasi-energy spectrum as the potential strength is changed in Fig.~\ref{fig:sinspecexamples} for two different system sizes. In the same plots, we colour the eigenstates according to their respective IPR defined in Eq.~\eqref{eqn:IPR}. 

In contrast to the static cases~\cite{Strkalj2022,Szabo2020} with $\omega=0$ shown in the insets of Fig.~\ref{fig:sinspecexamples}(a) and (b), we do not observe any sharp localization transitions of the majority of the spectrum. Most of the states remain delocalized, as indicated by the lower IPR values over the whole range of $W$ considered in Fig.~\ref{fig:sinspecexamples}. The localized states are sparse in the spectrum and appear only at unusually large $W$. We show this dynamical delocalization effect due to driving persists under scaling by studying several system sizes in Appendix.~\ref{sec:scaling}.
Note that the spectrum is particle-hole symmetric, which is not necessarily true for the static cases~\cite{Szabo2020, Strkalj2022}. In addition, in Fig.~\ref{fig:sinspecexamples}(a), we observe a narrowing of the spectral bandwidth when $W$ is increased, which is in stark contrast to the static cases where the energy spectrum widens for larger $W$. The narrowing persists for larger lattices shown in Fig.~\ref{fig:sinspecexamples}(b), although at a smaller rate compared to smaller systems.
Furthermore, a fraction of states in the quasienergy spectrum oscillate around zero, see Fig.~\ref{fig:sinspecexamples}(a). 

The narrowing of the bandwidth is a consequence of a dynamic localization effect albeit with a non-trivial structure due to the spatially varying multi-frequency nature of the driving~\cite{Dunlap1986,Lignier2007}. This follows from expressions in Eqs.~\eqref{eqn:phasetime} and~\eqref{eqn:alphas}, which we can simplify by taking the time average as the first term in the Magnus expansion~\cite{BlanesMagnus2009}, and find that the lowest order contribution in $W/\omega$ to nearest neighbour hopping amplitude is given by 
\begin{equation}
    J_{\mathbf{n} \mathbf{n}'}^{\text{eff}} \approx J \mathcal{J}_0 \bigg( \frac{W \alpha_{\mathbf{n} \mathbf{n}'}}{\omega} \bigg),
    \label{eqn:alphaforJ}
\end{equation}
with $\alpha_{\mathbf{n} \mathbf{n}'}$ being the dominant coefficient for the difference in driving between sites $\mathbf{n}$ and $\mathbf{n}'$. 
For a monochromatic and uniform driving, $\alpha=1$ in the expression above; all the hopping amplitudes in the system get renormalized by the Bessel function depending solely on the ratio $W/\omega$~\cite{Eckardt2017}. In this conventional case, the effective hopping amplitude decreases with increasing $W$ initially for constant $\omega$, reducing the bandwidth of the spectrum. 
However, in our case of non-uniform multi-frequency driving, one has to consider the fact that the driving frequency $\omega$ has to be rescaled by a spatially dependent factor since sites far from the centre locally experience a driving spectrum with higher frequencies, see Fig.~\ref{fig:drivingex}.

Although not reproducing fine details, the expression~\eqref{eqn:alphaforJ} does capture the reduction of bandwidth as a function of $W$ observed in Fig.~\ref{fig:sinspecexamples}. 
An interesting characteristics of periodic twisting is that this reduction is subdued for larger systems, as $\alpha_{\mathbf{n} \mathbf{n}'} \sim (\beta R)^{-1/2}$ decreases between sites further away from the axis of rotation, slowing down the reduction of $J_{\mathbf{n} \mathbf{n}'}^{\text{eff}}$. 
Surprisingly, while Eq.~\eqref{eqn:alphaforJ} strictly holds only in the limit $W/\omega \ll 1$, its range of applicability appears to extend beyond this as can be seen in Fig.~\ref{fig:sinspecexamples}. 

\begin{figure}[htbp]
  \centering
  \includegraphics[width=\linewidth]{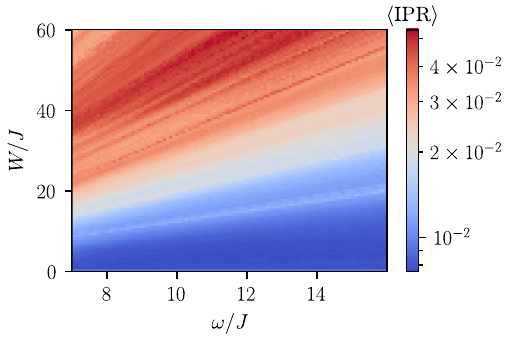}
  \caption{The averaged IPR over all the eigenvectors in the spectrum, $\langle\text{IPR} \rangle = \sum_m \text{IPR}^m/N^2 $, for $\Theta=\pi/2$ as a function of the potential strength and rotation frequency. Here we used a square lattice with a width $N=16$.
  The linear contours with fixed $W/\omega$ confirms the scaling predicted by Eq.~\eqref{eqn:alphaforJ}.}
  \label{fig:avglocsinsin}
\end{figure}

Another key prediction of Eq.~\eqref{eqn:alphaforJ} is that we should get similar localization properties for different values of $W$ and $\omega$ as long as their ratio $W/\omega$ remains unchanged. 
This is indeed the case, as we show in the averaged IPR across the spectrum, in Fig.~\ref{fig:avglocsinsin}, where clear lines with constant IPR for fixed $W/\omega$ are visible.
However, the localization does not just monotonically increase with increasing $W$; at sufficiently high values of around $W/J \approx 50$ and low frequencies, the oscillatory nature of the Bessel function starts appearing and the states become relatively delocalized.

We emphasize that the values of the averaged IPR are relatively low in Fig.~\ref{fig:avglocsinsin}, which is a consequence of majority of eigenstates being delocalized despite extremely high values of the perturbing potential strength (cf.~Fig.~\ref{fig:sinspecexamples}). Increasing the system size while keeping constant $W/\omega$ further reduces the average IPR since more delocalised states are introduced in the spectrum, which can be observed by comparing Figs.~\ref{fig:sinspecexamples}(a) and (b), in agreement with predictions of Eq.~\eqref{eq:RMS}. However, it is important to note that this does not mean that all states are delocalised, but there are states with high IPR, indicating localization, that are present in the spectrum for any system size and large enough $W$.

\begin{figure}
    \centering
    \includegraphics[width=1\linewidth]{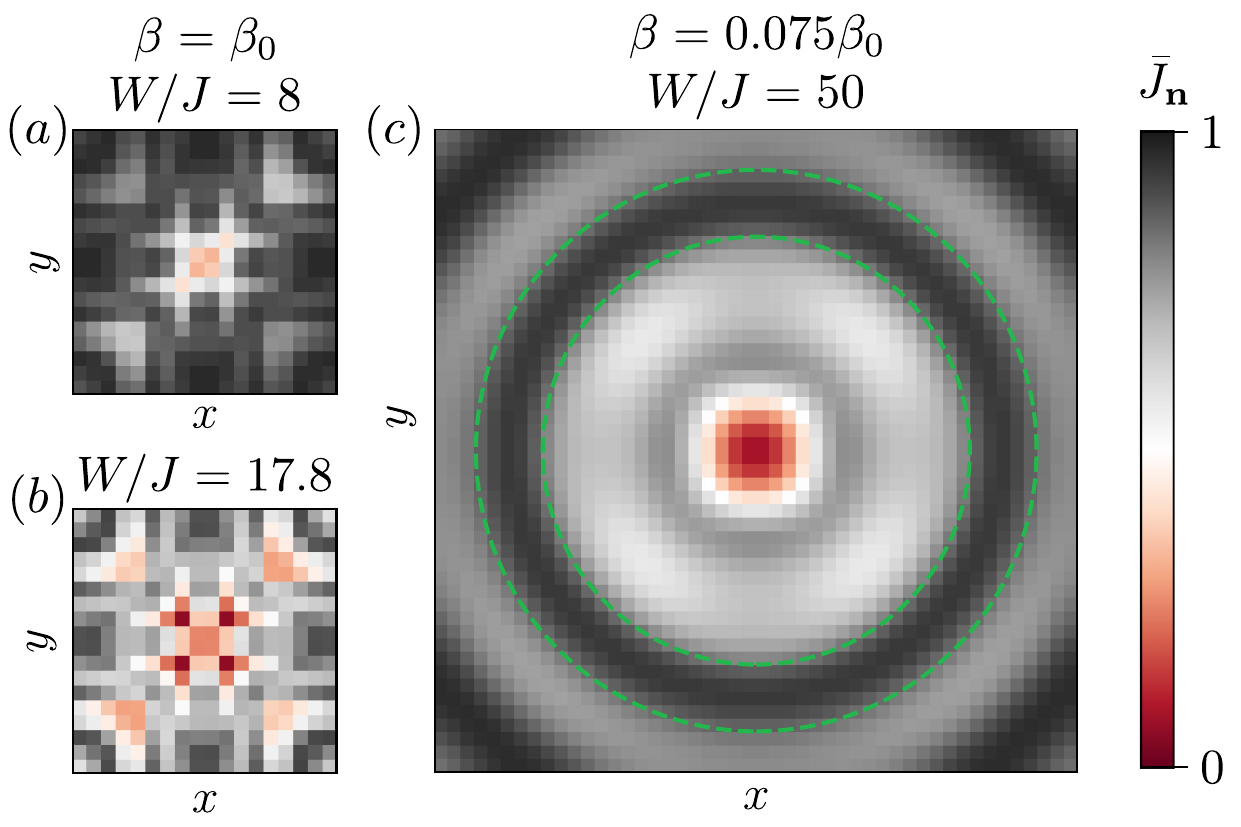}
    \caption{The nearest-neighbour hopping amplitudes averaged over all four directions $(\overline{J}_{\mathbf{n}})$ on a site, $\mathbf{n}=(x,y)$, as defined in Eq~\eqref{eq:spatially_averaged_hopping_amplitude}. Numerical results for a square lattice of width $N=18$ and $\beta=\beta_0 = 2/(1+\sqrt{5})$, with different potential strengths (a) $W/J=8$ and (b) $W/J=17.8$.
     (c) $\overline{J}_{\mathbf{n}}$ for a lattice with $N=48,\,\beta = 0.075 \beta_0$ and $W/J=50$. Green dashed concentric circles bound the area with averaged hopping amplitudes close to 1. In all plots, we used $\omega/J=9$.}
    \label{fig:avgJ}
\end{figure}

{\it Spatial distribution of the effective tunnelling}--- Alongside describing the spectral bandwidth as a function of $W/\omega$ in Fig.~\ref{fig:sinspecexamples}, the effective tunnelling amplitude in Eq.~\eqref{eqn:alphaforJ} signals a richer spatial interplay for the radial profile that can emerge under periodic twisting. To analyze this, we calculate the average tunneling for each site $\mathbf{n}$ over all four bonds that connect it to its nearest neighbours as
\begin{align}   \label{eq:spatially_averaged_hopping_amplitude}
    \overline{J}_{\mathbf{n}} = \frac{1}{2} \sqrt{\bigg(\sum_{\mathbf{n}'=\pm \hat{x}} J_{\mathbf{n},\mathbf{n}+\mathbf{n}'} \bigg)^2 + \bigg(\sum_{\mathbf{n}'=\pm \hat{y}} J_{\mathbf{n},\mathbf{n}+\mathbf{n}'}\bigg)^2}, 
\end{align}
where $x$ and $y$ are directions in quadrature, and the values of $J_{\mathbf{n},\mathbf{n}+\mathbf{n}'}$ are taken from the numerically calculated Floquet Hamiltonian~\eqref{eq:U}.
We present the numerical results in Fig.~\ref{fig:avgJ}.

For a wide range of $W/J$, there is a region spanning several sites in the centre of the lattice with low hopping amplitudes; see Figs.~\ref{fig:avgJ}(a) and (b). This separates the central region of the lattice, where the states gain less kinetic energy due to weaker tunnelling, from the outer region, where they acquire more kinetic energy. 
In Fig.~\ref{fig:DOSplots2}, this is nicely observed in the spatial distribution of the eigenstates, where a distinction is visible between the highly localized central region of roughly $4 \times 4$ sites and the surrounding more delocalized parts at sufficiently large $W$.

The origin of these highly suppressed tunneling amplitudes on average near the axis of rotation can be understood from Eq.~\eqref{eqn:alphas}, indicating a stronger Floquet renormalization of tunneling. Near the centre, $\alpha_{\mathbf{n} \mathbf{n}'}$ is large since the value of the coefficients given by the Bessel function is larger for smaller $\gamma_\mathbf{n}$. Hence, larger $\alpha_{\mathbf{n} \mathbf{n}'}$ in Eq.~\eqref{eqn:alphaforJ} gives rise to smaller effective hopping amplitude, successfully capturing the more localized centre. 

Moreover, Eq.~\eqref{eqn:alphaforJ} predicts Bessel-like oscillations which become even clearer in larger systems and smaller quasiperiodic frequency $\beta$, as can be seen qualitatively for the average hopping amplitude $\overline{J}_{\mathbf{n}}$ in Fig.~\ref{fig:avgJ}(c). While the spectral content of the drive (cf.~Fig.~\ref{fig:drivingex}) expands at larger distances, Fig.~\ref{fig:avgJ}(c) highlights the fact that the multi-frequency nature of periodic twisting emerges in a non-linear way. 

Indeed, the average hopping amplitudes remain large in the circular regions shown in Fig.~\ref{fig:avgJ}(c) even as the strength of the potential is greatly enhanced, drawing a stark contrast with static settings~\cite{Strkalj2022,Szabo2020}. Here, shown for large $W/J=50$, strong enough to suppress tunneling at equilibrium, we observe that the periodic twisting dynamically induces several effective weak-potential lines within which transport should remain feasible. 

{\it Fractal dimension}--- We expect the peculiar spatial structure of the hopping amplitudes to reflect itself in the localization properties of particular eigenstates.
Therefore, to characterise the overall localization behaviour of the eigenstates further, we examine the fractal dimension of each eigenstate (marked by $m$) given by
\begin{equation}
    D_m(q) = \frac{1}{1-q} \frac{\ln\left(\sum_{\mathbf{n}} |\psi^m_{\mathbf{n}}|^{2q}\right)}{\ln(N^2)} \, ,
\end{equation} 
where the sum goes over every site $\mathbf{n}$ on the lattice, with a total of $N^2$ sites~\cite{Evers2008,Fracdim}. 
While the localized states correspond to $D_m(q) =0$, the ones extended over the whole system give $D_m(q) =1$ in the thermodynamic limit~\cite{Evers2008}.  
\begin{figure}
    \centering
    \includegraphics[width=1\linewidth]{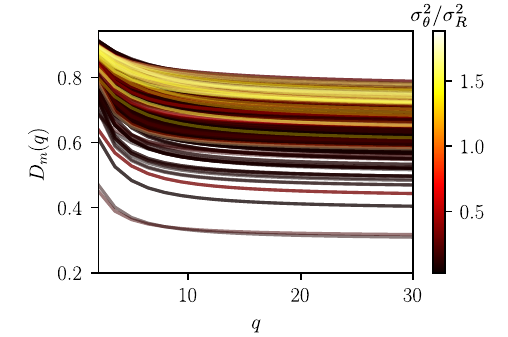}
    \caption{Fractal dimension for quasienergy eigenstates $\psi^m$ for $\omega/J=9$, $W/J=50$ and $\beta=0.15\beta_0$ for lattice size $N=24$.
    The colorbar shows the ratio between the variances of the eigenstate's density in the angular and radial directions as defined in the main text. This measure emphasises the states localized on a ring region with a large spread in $\theta$ and a small spread in $R$.}
    \label{fig:fracdim}
\end{figure}

We present the numerically calculated $D_m(q)$ in Fig.~\ref{fig:fracdim} and first discuss its overall trend here. 
The diverse behaviour of the fractal dimensions for different states pinpoints the non-uniform behaviour of localization properties throughout the spectrum, inline with the discussion above. Several states have fairly large $D_m$, indicating their extended nature. These show some $q$ dependence, which is expected even for fully extended states in finite-size systems. Interestingly, despite the high strength of the potential relative to the bare tunnelling, a significant fraction of states remain delocalized also at this smaller value $\beta=0.15\beta_0$.
Alongside the aforementioned extended states, there are also a few states with low fractal dimensions. Those states with small $D_m$ values are localized on the central sites, closest to the axis of rotation, as we discussed above. 
The dependence on $q$ of the lowest few states in Fig.~\ref{fig:fracdim} is also attributed to the finite size of the system. We relegate further results on the fractal dimension for different system sizes to Appendix.~\ref{sec:scaling}.

{\it Radial measure of localization}--- While the fractal dimension $D_m$ is a useful tool to distinguish exponentially localized states, with decay in all directions, from critical or extended ones, it fails to give a deeper insight into more exotic types of localization, e.g.~localization inside a ring-shaped region. Suggested by the spatial profile of the averaged hopping amplitudes $\overline{J}_{\mathbf{n}}$ in Fig.~\ref{fig:avgJ}, we expect that some states might be confined inside the dashed region where $\overline{J}_{\mathbf{n}}$ is of the order of unity. 
To obtain information on the degree of radial confinement of eigenstates, we thus introduce a radial measure of localization as the ratio of the spatial density of each eigenstate spreading in angular and radial directions. More precisely, we look at the ratio between the variances associated with the probability density $|\psi^m|^2$ in angular and radial directions, $\sigma^2_{\theta}/\sigma^2_{R}$. We define these variances explicitly in Appendix.~\ref{sec:scaling}. When this ratio is large, the state is more confined in radial than in angular direction, in line with localization within a ring. 
For these ring-localized states, we expect $D_m$ to have a relatively high value, since the ring occupies a larger part of the lattice --compared to conventional, strong exponential localization along all directions that occurs in the undriven case~\cite{Szabo2020,Strkalj2022}-- but simultaneously having a large ratio of $\sigma^2_{\theta}/\sigma^2_{R}$, which excludes the possibility of the state being extended over the whole system.

In Fig.~\ref{fig:fracdim}, we color each line (eigenstate) with its respective $\sigma^2_{\theta}/\sigma^2_{R}$. Interestingly, we find many states that fit the description of ring-localized states defined above. For example, the bundle of states located at relatively high $D_m(q=2) \approx 0.8$ values at the same time has large $\sigma^2_{\theta}/\sigma^2_{R}$ ratios (colored light yellow). We note that although $\sigma^2_{\theta}/\sigma^2_R \approx 2$ appears small, since $\sigma^2_{\theta}$ has a maximum value of 1, this corresponds to a radial standard deviation $\sigma_R < 1/\sqrt{2}$, which is less than a single site. To further support these claims, we study the fractal dimension scaling with system size in Appendix.~\ref{sec:scaling} and show strong evidence for robust localization of the rings for increasing system size $N$ ($D_m(q)\to0$). Although they are formally localized in the thermodynamic limit, their quasi-one-dimensional nature allows these ring states to exhibit interesting spectral properties, transport signatures and local topological markers as we address next.

\section{Ring States} \label{Sec:RingState}
\begin{figure}
    \centering
    \includegraphics[width=\linewidth]{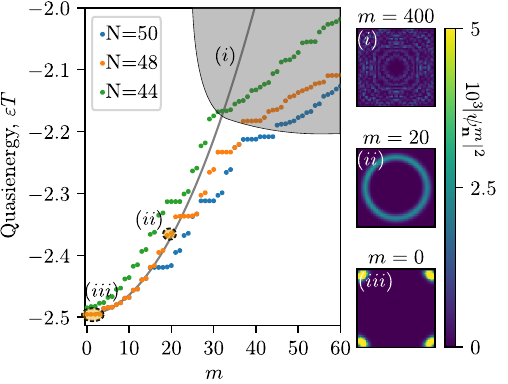}
    \caption{Spectra for $W/J=50$ and $\beta=0.015\beta_0$ for different system sizes $N$ demonstrating the ring (ii) and pocket (iii) states at low quasienergies. The parabola illustrates the ring-like character of the states, as their energies scale like $\propto m^2$. Inset (i) shows a generic bulk state with its density distributed across the whole system; (ii) a ring state with almost two-fold degeneracy; (iii) a pocket state with almost four-fold degeneracy marked on the main spectra.}
    \label{fig:ring_states_spectrum}
\end{figure}
The appearance of the quasi-one-dimensional regions harbouring ring-shaped extended states, which we dub ring states, is one of the key features emerging under periodic twisting. 
Since the average hopping strength $\overline{J}_{\mathbf{n}}$ is higher inside the ring, as alluded to in Fig.~\ref{fig:avgJ}, we expect the ring states to have larger absolute values of quasienergy and, consequently, live close to the edge of the spectrum. 
Hence, we show the lower end of the spectrum in Fig.~\ref{fig:ring_states_spectrum} as a function of the state number $m$ for different system sizes. There, we observe three major characteristics: (i) the states with smaller $|\varepsilon T|$ are in general extended over the whole system, (ii) at higher values of $|\varepsilon T|$, well separated from the bulk states, there are almost degenerate pairs of states whose quasienergies follow a parabola (solid line), and (iii) there are degenerate quadruplets of states that are scattered towards the ends of the spectrum. 

The extended states marked with (i) strongly depend on system size. Both their energies and the spatial profile of their density change with increasing the linear size $N$, as can be seen in the density of such a state given in the inset. We refer to them as bulk states. 

In contrast, we find that the states marked with (ii) do not show any system-size dependence; their energies are identical for all three sizes shown in the left panel of Fig.~\ref{fig:ring_states_spectrum}. 
We call these states ring states, following from their ring-shaped density distribution given on the right-hand side panel in Fig.~\ref{fig:ring_states_spectrum}. 
The spatial extent of the density of these states coincides with the region of large $\overline{J}_{\mathbf{n}}$ bounded by green dashed lines in Fig.~\ref{fig:avgJ}.
Furthermore, ring states feature a characteristic parabolic dependence on state number $m$ and a pairwise (almost) degeneracy, which can be understood from a simpler model of a one-dimensional chain with periodic boundary conditions. The momentum in a chain of length $L$ is discretized as $ k_m=2\pi m/L$, leading to discrete energies of $E_m = -2 J_{\rm eff} \cos(k_m a) \approx -2 J_{\rm eff} (1 - 2 \pi^2 m^2 /L^2)$, where $a$ is the distance between two neighbouring sites in the chain and $J_{\rm eff}$ is a constant hopping amplitude. The boundary conditions for the wave function and its derivative lead to pairwise degenerate states in energy, corresponding to different angular momentum modes along the ring.  
We emphasize that although this one-dimensional model correctly captures the parabolic energy dispersion, a direct fit to numerics is more involved as it requires estimating the prefactor $J_{\rm eff}$ from the tunnelling strengths in the Floquet Hamiltonian, which is impractical due to the non-uniform behaviour of averaged hopping amplitudes within the ring region that is only quasi one dimensional. We also note that the lifted degeneracy of the pairs is not captured by the simplified model. This originates from the time-reversal symmetry breaking in the system, which we discuss in detail in Sec.~\ref{SecSub:RingTopology}.

Lastly, the states marked with (iii) in Fig.~\ref{fig:ring_states_spectrum} are localized in corners of the system and appear due to a finite size. The average hopping strengths $\overline{J}_{\mathbf{n}}$ oscillate as one moves away from the axis of rotation, forming successive dynamically-induced weak-potential lines, and the tunnelling becomes stronger again within a third ring-shaped region as can be seen in Fig.~\ref{fig:avgJ}(c). The finite square geometry of the system cuts this third ring of high $\overline{J}_{\mathbf{n}}$, leading to four equal spatially-disconnected pockets close to the corners of the square lattice. Similarly to (ii), the states living in these pockets will have energies that reside at the ends of the spectrum, meaning that the hybridization with the bulk states is absent and the states are localized within the corner pockets. However, since the pockets are disconnected, unlike the ring states, they are four-fold degenerate and do not follow a parabolic dispersion. 
Crucially, changing the system size alters the size of the pockets and leads to different energies of quadruplets for different $N$, see Fig.~\ref{fig:ring_states_spectrum}, which would eventually connect to form a full ring in larger systems with the quadruplets splitting into pairs of states.

We note that the size and the number of delocalized ring regions can be controlled by changing the linear size $N$ or tuning the spatial frequency $\beta$ of the perturbing lattice. We show such a configuration where the third ring fully fits into the system in Appendix~\ref{sec:ring_hybridization} with the fourth one also visible. 
Once multiple ring regions of large $\overline{J}_{\mathbf{n}}$ are present in the system, the ring states residing on different rings form independent parabolic dispersions, due to different circumferences of the delocalized regions. Separation by several sites in the radial direction leads to weak coupling between different rings, and their states are only sparsely hybridized with each other throughout the spectrum. This hybridization occurs only when the difference in energy of states living in different parabolas closely matches, such that the effective hybridization element is comparable to the energy difference. Therefore, we find that many of the ring states are strongly confined within their respective annulus regions, leading to the physics discussed above, even in much larger systems.
Note, however, that this is valid only for the states that are not in the bulk of the spectrum, but reside on the band edges. For the ring states whose quasienergies fall in the bulk of the spectrum, see the grey area in Fig.~\ref{fig:ring_states_spectrum}, the hybridization among different rings is enhanced by the bulk states that extend over the whole system, weakening the strong ring confinement.

\begin{figure}[t]
    \centering
    \includegraphics[width=1\linewidth]{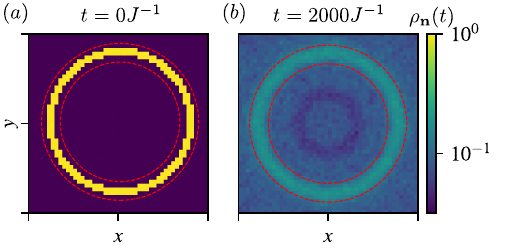}
    \caption{Transport dynamics associated with the ring states, for the evolution of a wave packet, $\Psi_i(\mathbf{n},0)=\delta_{\mathbf{n},\mathbf{n}_i}$, initialized on a single site, $\mathbf{n}_i$. We show the total density distribution $\rho_\mathbf{n}(t)= \sum_{i=1} |\Psi_i(\mathbf{n},t)|^2$ of several wave packets at time (a) $t=0 J^{-1}$ and (b) $t=2000 J^{-1}$. While some small portion escapes, the majority of the density remains within the ring marked with dashed lines. Parameters are same as Fig.~\ref{fig:avgJ}(c).}
    \label{fig:48transp}
\end{figure}

{\it Transport signatures}--- The number of ring states residing in a particular annulus region of high average hopping amplitude is not extensive and is not affected by increasing the system size. To increase the number of ring states in each ring, one needs to change the spatial frequency $\beta$ instead, which consequently changes the radius of the ring. 
That being said, one can ask whether such a small, non-extensive number of ring states can still give rise to experimentally observable signatures. 
To demonstrate this, we investigate the transport dynamics of a wave packet~\cite{Martinez23_wavepacket} initialized on the larger ring marked with dashed lines in Fig.~\ref{fig:avgJ}(c). To ensure that our procedure does not select any special sites within the ring, we initialize the wave packet on different sites and present the averaged dynamics in Fig.~\ref{fig:48transp}. Note that the evolution of the wave packages are mutually independent. More results on a single wave packet transport are detailed in Appendix~\ref{sec:ring_transport_24x24}.

We show the numerical results for the probability density $\rho(t)$ in Fig.~\ref{fig:48transp}, where we depict the starting sites and  the long-time dynamics at $t=2000 J^{-1}$. We observe that the majority of $\rho(t)$ remains confined within the ring region marked with the dashed lines even after several thousand hopping times, with a minor leakage to other parts of the system due to the small overlap with bulk states that also extend inside the ring. 
Surprisingly, there is a density depletion within the area coinciding with the smaller ring shown in Fig.~\ref{fig:avgJ}(c). This is due to the fact that ring states residing in different rings have minimal overlap, and  hence, support a stable transport localized within the bulk. 
These numerical results align with the conclusions of the previous analyses, highlighting significant potential for experimental observations.

\section{Topological Signatures}
Topologically protected states are known to survive in spatially inhomogeneous, disordered~\cite{Bellissard1994, Hastings+Loring2011, Prodan2010} or aperiodic systems in static or dynamically modulated settings~\cite{Bandres16_PRX,Rosa2021,Johnstone2022,Titum2016_PRX_AFAI, Nathan2019_PRB_AFI,BaiWeld2025}. 
Notable examples include fractal topological gaps emerging in a quasicrystal~\cite{Bandres16_PRX} and higher-dimensional topological physics emanating from sliding twisted layers~\cite{Rosa2021}. 
The cases investigated in the literature have mostly concerned aperiodic systems under a constant magnetic field~\cite{Hatakeyama89_JPS,Tran15_PRB,Fuchs18_PRB,Huang18_PRB_QSH} which in general bring the Hofstadter physics~\cite{Hofstadter1976} into quasicrystalline settings. These considerations have also been found to yield counterintuitive states with non-trivial topological invariants, which are localized in the bulk of a quasicrystal~\cite{Johnstone2022,BallingNielsen_26BLT}, hence dubbed bulk-localized states, and attracted attention as their topological origin is not fully understood. 

The periodic twisting introduced in this work, which does not trivially localize all eigenstates as evidenced by delocalized rings even under strong potentials, 
offers an interesting resemblance with these topologically protected states and may provide an original tool to explore topological physics. 
Motivated by the rich topological phenomena in aperiodic systems~\cite{Prodan2010,Bandres16_PRX,Rosa2021,Johnstone2022}, some of which have no crystalline counterparts, we here investigate whether periodic twisting gives rise to any topological signatures by calculating real-space topological markers. Interestingly, we find that some states in the spectrum carry non-trivial Bott indices and Chern markers and reside in the dynamically-induced delocalized ring regions. We analyse the source of these signatures and identify them to be originating from the weak time reversal symmetry (TRS) breaking and hybridization of the different delocalized regions, whose protection, however, remains weak due to small gaps in the spectrum. 
We present an analysis contrasting with Chern insulators in disordered limits. Below, we introduce the real-space topological invariants and explore their features in a wide range of parameters, which goes beyond clean quasicrystal cases under a uniform magnetic field that have been the focus of attention so far.

\subsection{Complex Tunnelling Amplitudes}
\label{sec:ComplexNNN}
\begin{figure}
    \centering
    \includegraphics[width=\linewidth]{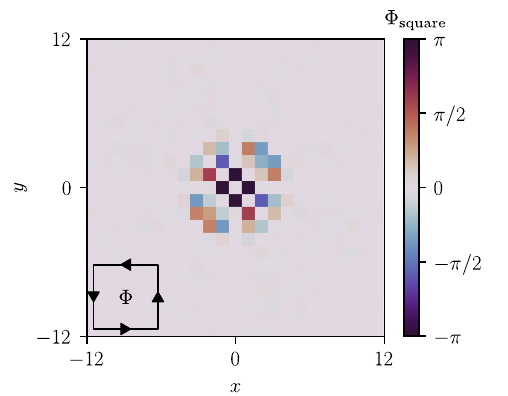}
    \caption{Effective magnetic flux $\Phi_\text{square}$ piercing through each square plaquette captured by Peierls phases in Eq.~\eqref{eqn:fluxdef}, as illustrated in the inset. Here $W/J=17.8$, $\omega/J=9$ and $\beta=\beta_0$ on a square lattice of width $N=24$. The lattice sites sit on the vertices of the squares shown in the plot, with the inset depicting the flux. The antisymmetry of the flux distribution along the diagonal stems from the antisymmetry of the underlying driving potential (cf.~Fig.~\ref{fig:potentialexample}(c)).     }
    \label{fig:fluxdistex}
\end{figure}

Before introducing topological markers and investigating the spectra, it is instructive to examine the Hamiltonian matrix elements instigated by periodic twisting to motivate this analysis. 
The point is that the rotational nature of periodic twisting can form a ground for breaking TRS that is necessary for inducing topologically protected states in our two-dimensional setting~\cite{HaldaneOriginal}, although in a complicated way due to the spatially-inhomogeneous and local multi-frequency character of the drive. 
The broken TRS ultimately manifests itself as complex phases for the tunneling amplitudes, which can give rise to non-zero effective magnetic fluxes in the system. To see this, we inspect the effective tunneling parameters $J^F_{\mathbf{n}\mathbf{n}'}$ in the numerically calculated Floquet Hamiltonian~\eqref{eq:U} and extract the Peierls phases, defined as the argument $\phi_{\mathbf{n}\mathbf{n}'}=\text{arg}(J^F_{\mathbf{n}\mathbf{n}'})$. The directional sum of these Peierls phases upon circling around a closed loop is gauge invariant,
\begin{equation}
    \prod_{\mathbf{n},\mathbf{n}' \in \text{Plaquette}} e^{i\phi_{\mathbf{n}\mathbf{n}'}} = e^{i\Phi},
    \label{eqn:fluxdef}
\end{equation}
and corresponds to the effective magnetic flux piercing through this loop, where we take the product moving counter-clockwise around it (depicted in Fig.~\ref{fig:fluxdistex} inset). 
Importantly, while the nearest-neighbor hopping amplitudes in the primary square lattice are all real, we indeed find that periodic twisting can induce local magnetic fluxes different from zero or $\pi$ as shown in Fig.~\ref{fig:fluxdistex}. 
We note that although the values for the effective flux distribution depend on the Floquet gauge~\cite{Nixon24_QST}, we always obtain non-zero fluxes in the system.

Even though the flux distribution is highly non-uniform, stemming from the underlying local inhomogeneous driving, we generally observe significant fluxes around the central region when only the nearest-neighbor terms are considered and $W$ is small. It is worth pointing out that the tunneling gets strongly suppressed as well at the very centre of the system (cf.~Fig.~\ref{fig:avgJ}). Periodic twisting, however, can also promote tunneling between next-nearest neighbour sites along the diagonals as indicated by the Magnus expansion~\cite{BlanesMagnus2009}. While these higher-order processes are negligible when the driving is weak, they can become significant for strong driving or when the nearest neighbour hopping is suppressed.

\begin{figure}
    \centering
    \includegraphics[width=1\linewidth]{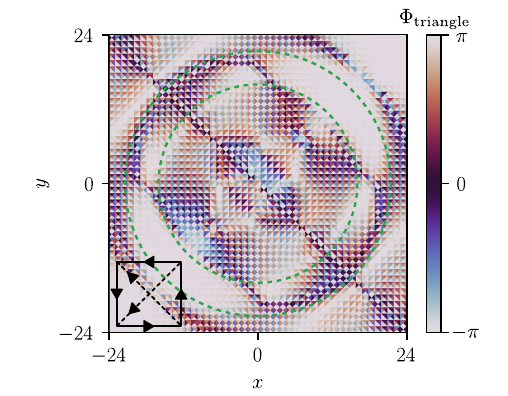}
    \caption{The flux distribution through triangle quarters within each plaquette originating from complex next-nearest neighbour tunnelling amplitudes (see inset and the main text for definition). For the chosen values of strong driving with $W/J=50$, $\omega/J=9$ and $\beta=0.075\beta_0$, the total flux through each square plaquette is zero to a good approximation. 
    Green concentric circles mark the region where averaged nearest-neighbour hopping amplitudes remain large in Fig.~\ref{fig:avgJ}(c).     }
    \label{fig:nnntunnellingphasemag}
\end{figure}

For strong disorder $W/J =50$, we find that the total flux through each square plaquette ($\Phi_\text{square}$) becomes negligibly small, in the order of $10^{-2}$, across the whole lattice. However, considering next-nearest neighbour tunneling processes in the Floquet Hamiltonian, we can study the local flux within a plaquette, accumulated in the triangular quarters ($\Phi_\text{triangle}$) of the square tiles as illustrated in the inset of Fig.~\ref{fig:nnntunnellingphasemag}. 
Note that this is similar to the non-uniform flux distribution in the Haldane model~\cite{HaldaneOriginal}, which in our case varies also from plaquette to plaquette.
While there is some inherent gauge freedom in $\Phi_\text{triangle}$ as in the Haldane model, we demonstrate a distribution of these local fluxes emerging from next-nearest neighbour tunnellings in Fig.~\ref{fig:nnntunnellingphasemag}, see Appendix \ref{sec:evensymloc} for details.  
Interestingly, we find that there are some significant flux values in the lattice and the TRS is broken locally and inhomogeneously.  
The region where we observe ring localization, with large average nearest-neighbor hopping strengths, see Fig.~\ref{fig:avgJ}(c), is marked with green concentric rings. The next-nearest neighbour tunnelling processes attain also non-negligible amplitudes within this area, and we find that the delocalized ring region harbours flux values that can get considerably large locally, which can give rise to topologically non-trivial signatures as will be discussed below.

\subsection{Topological Invariants in Real Space}
Having demonstrated the presence of effective local magnetic fluxes in the system induced by periodic twisting, we now turn to topological invariants defined in real space. 
Due to the spatial non-uniformity of our 2D model, it is not possible to directly calculate a Chern number in the absence of translational invariance, similar to the cases of a quasicrystal, systems with disorder or open boundary conditions where a periodic Brillouin zone cannot be defined~\cite{Hastings2010,Hastings+Loring2011,bianco2014chern, loring2019guide,Toniolo2022}. 
Therefore, we here employ the Bott index and Chern marker which allow for calculating the topological invariant directly in real space.  
The Bott index $\mathcal{B}^m$ for the $m$th state in the spectrum is defined as
\begin{equation}
    \mathcal{B}^m = \frac{1}{2\pi} \operatorname{Im} \{ \operatorname{Tr} \big[ \log(\hat{V}_x^m \hat{V}_y^m \hat{V}_x^{m \dagger} \hat{V}_y^{m \dagger}) \big] \},
    \label{eqn:Bottdef}
\end{equation}
and distinguishes whether $\hat{V}_{x} $ and $\hat{V}_{y}$ can be in principle approximated as commuting matrices or not~\cite{Hastings2010}. Here, these matrices 
\begin{align}
    \hat{V}^m_{x} &= \hat{Q}^m + \hat{P}^m \hat{U}_{x} \hat{P}^m, \nonumber\\
    \hat{V}^m_{y} &= \hat{Q}^m + \hat{P}^m \hat{U}_{y} \hat{P}^m,
\end{align}
correspond to two orthogonal directions of the two-dimensional system where  $\hat{U}_{x} = \exp(2 \pi i \hat{x}_\mathbf{n} )$ and $\hat{U}_{y} = \exp(2 \pi i \hat{y}_\mathbf{n} ) $ for the position operators $\hat{x}_\mathbf{n}$ and $\hat{y}_\mathbf{n}$ scaled between 0 and 1, and $\hat{P}^m = \sum_{i\leq m} \ket{\psi^i} \bra{\psi^i} = \hat{\mathbb{1}}-\hat{Q}^m$ is the Fermi projection. 
While the Bott index generally vanishes in a topologically trivial system, it takes finite and quantized values for in-gap states, which corresponds to the Chern number~\cite{Toniolo2022} of the occupied states below and signals an edge state at the given energy. Essentially, by revealing the non-commutativity of $\hat{V}_{x} $ and $\hat{V}_{y}$, a non-zero Bott index indicates a topological obstruction to find exponentially localized Wannier functions along both directions for the projected bands~\cite{Hastings+Loring2011,Hastings2010}, as exactly captured by the Chern number~\cite{Vanderbilt2018}. This brings the clear advantage that it can be calculated numerically in real space surpassing the need for a band structure in the momentum space, which makes it suitable for our setting.

Another powerful invariant that can be calculated locally in real space is called the Chern marker by extending the definition of the Chern number in a set of Bloch bands to position space~\cite{BiancoResta2011,kitaev2006anyons}. It is defined on each lattice site $\mathbf{n}$ as
\begin{equation}
    \mathcal{C}_\mathbf{n}^m = -\frac{4\pi}{A_c} \operatorname{Im}\{ \bra{\mathbf{n}} \hat{x}^m \hat{y}^m \ket{\mathbf{n}} \},
    \label{eqn:Cmarker}
\end{equation}
where we again consider the gap between the $m$th and $(m+1)$th eigenstates and use the projected position operators $\hat{x}^m = \hat{Q}^m \hat{x} \hat{P}^m$ and $\hat{y}^m =  \hat{P}^m\hat{y}\hat{Q}^m$. The parameter $A_c$ is the size of the unit cell, which is scaled to 1 in our case. 
For open boundary conditions, the Chern marker for a given eigenstate sums up to zero over the entire system ($\sum_\mathbf{n} \mathcal{C}_\mathbf{n}^m = 0$) as the global Hall conductivity vanishes in a finite open system. Notably, the Chern marker can take stable and quantized values across extended regions of the lattice, signalling underlying non-trivial Chern numbers~\cite{BiancoResta2011}. The interplay with intricate real-space structures can even generate spatially alternating regions of different local invariants forming Chern mosaics in moir\'e settings or quasicrystals~\cite{Antao2025_ChernMosaicTN,Grover22_NatPhys_ChernMosaic}.

\subsection{Topological Signatures of Ring States}  \label{SecSub:RingTopology}

\begin{figure}
    \centering
    \includegraphics[width=1\linewidth]{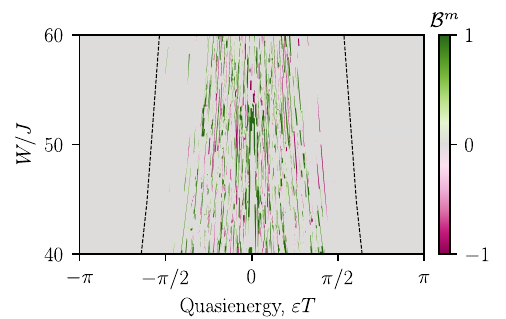}
    \caption{Bott indices across the quasienergy spectrum as the potential strength is varied. The bandwidth (marked with dashed lines) becomes narrower in agreement with Eq.~\ref{eqn:alphaforJ}. Finite Bott signatures are obtained for a range of $W$, demonstrating some robustness. Parameters are $\beta=0.15\beta_0$, $\omega=9$ and $N=24$.}
    \label{fig:bottWvary}
\end{figure}

We calculate the Bott index and the Chern number numerically by using the Floquet eigenstates to analyse how topologically non-trivial signatures can emerge under periodic twisting. The Bott index becomes finite for several groups of states across a range of parameters as demonstrated in Fig.~\ref{fig:bottWvary} as a function of the driving strength $W$. 
While there is a zoo of states with non-trivial Bott invariants, we find that the localization behaviour, which isolates the central region of the lattice from the outside while giving rise to the ring states (cf.~Fig.~\ref{fig:avgJ}), also has a strong effect on the topological signatures. 
Indeed, some of these states marked in Fig.~\ref{fig:bottWvary} are found to have a distinctive circular shape. 

We here focus on these delocalized ring states with a non-trivial Bott index for which the Chern marker also becomes relatively uniform over extended regions of several lattice sites, and give examples of other states with non-trivial Bott indices in the appendix.  Fig.~\ref{fig:loc+Chernbig} shows the Chern marker for such a ring state. Despite the strong perturbing potential $W=50J$, this state features a Bott index of 1 and the Chern marker  saturates to the unit value within an annulus region marked by red dashed lines, which corresponds directly to the conductive ring in Fig.~\ref{fig:avgJ}. The state depicted in Fig.~\ref{fig:loc+Chernbig} indeed lives at the bottom end of the spectrum in line with the discussion in Sec.~\ref{Sec:RingState}.

\begin{figure}
    \centering
    \includegraphics[width=1\linewidth]{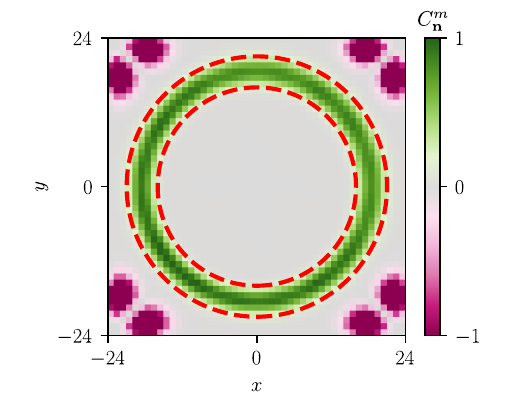}
    \caption{Chern marker, $C_\mathbf{n}^m$, of a ring state $m=21$ with the Bott index $\mathcal{B}^m=1$. Red dashed lines indicate the conductive ring with large average effective tunnelling amplitudes from Fig.~\ref{fig:avgJ}(c). The parameters are $W/J=50$, $\omega/J=9$ and $\beta=0.075\beta_0$ for a linear lattice size $N=48$.}
    \label{fig:loc+Chernbig}
\end{figure}

In order to identify the origin of the non-trivial topological signatures in ring states, we inspect their properties closer. 
Firstly, we find that the non-trivial Bott invariant arises from the weak TRS breaking under periodic twisting detected by non-zero local flux values as discussed in Sec.~\ref{sec:ComplexNNN}. 
To see this, we consider two different modifications for the complex next-nearest-neighbour (pairs labelled by ${\{\{\mathbf{n},\mathbf{n}'\}\}}$) tunnelling processes in the Floquet Hamiltonian; by setting their amplitude to zero $J^F_{\{\{\mathbf{n},\mathbf{n}'\}\}}=0$ and by setting their phase to zero~\footnote{We note that while there can be still a small phase associated to the nearest-neighbour tunnelling amplitudes, these become negligible in strong driving regimes.}. 
Note that the average nearest-neighbour tunnelling strengths still remain large, forming the quasi-one-dimensional delocalized regions as in Fig.~\ref{fig:avgJ}. In both cases, we indeed obtain vanishing Bott indices for the ring states, highlighting the critical role played by these complex hopping amplitudes induced by periodic twisting.

\begin{figure}
    \centering
    \includegraphics[width=\linewidth]{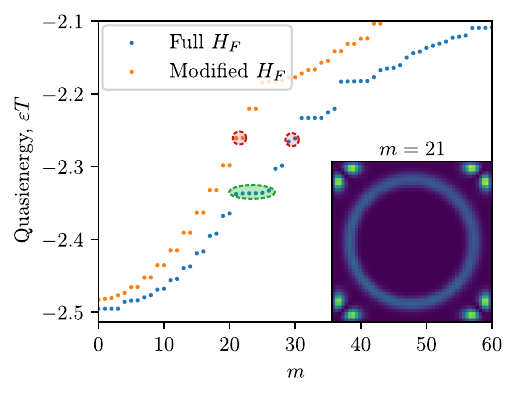}
    \caption{Comparison of the energy spectra with a modified $H_F$ where terms other than the nearest neighbour tunnellings are removed. The complex higher order terms in $H_F$ are associated with weak TRS breaking, resulting in weak degeneracy breaking of the paired ring states (circled in red), which is absent in modified $H_F$. The quadruple pocket states at the corners are also removed by imposing a circular wall. The non-trivial topological signatures of ring states (example density distribution given in the inset for the state shown in Fig.~\ref{fig:loc+Chernbig}) arise from the TRS breaking and hybridization between ring and pocket states (circled in green).    }
    \label{fig:RingStHybridz}
\end{figure}

The second key observation is that the finite Bott index associated with the ring states originates from hybridization between the delocalized regions in the lattice. 
For the system size considered in Fig.~\ref{fig:loc+Chernbig}, there are pocket states at the four corners (labelled $(iii)$ in Fig.~\ref{fig:ring_states_spectrum}) lying at the radius where the average tunnelling amplitudes become large once more (see~Fig.~\ref{fig:avgJ}). 
We find that non-trivial Bott values appear when the pairs of ring-localized states ( $(ii)$ in Fig.~\ref{fig:ring_states_spectrum}) hybridize with the pocket quartet.
Indeed, imposing a circular wall to cut the delocalized pockets out makes the quadruples disappear (see orange dots in Fig.~\ref{fig:RingStHybridz}) and yields a vanishing Bott index for ring states.
We note that this is not a mere finite size effect with the pocket states appearing due to the square geometry. The lattice size and $\beta$ can be tuned so that the third delocalized ring fully fits into the system as we show in Appendix~\ref{sec:ring_hybridization}. In this case, we again observe finite Bott values from the mixing of the second ring eigenstates with the third ring, underpinning the hybridization between different delocalized regions in the system as a key element for non-trivial invariants.

We demonstrate these findings in Fig.~\ref{fig:RingStHybridz}, with the inset showing the density profile of the eigenstate whose Chern marker distribution is given in Fig.~\ref{fig:loc+Chernbig}. This state has a non-trivial Bott index and lies at an energy where the paired states of the second ring hybridize with the quadruple pocket states (circled in green). We also plot the spectrum of a modified Hamiltonian which is confined within a circular geometry leaving the corner pockets out and includes only the nearest neighbour hopping terms. Due to the circular wall, the degenerate quartets disappear and the energies of the quasi-one-dimensional ring region follows a parabola as discussed in Sec.~\ref{Sec:RingState}. Moreover, in the absence of higher order terms, the pairs of ring states become degenerate (red circles)~\footnote{Note that we remove all further than nearest-neighbour couplings and on-site terms in the modified Hamiltonian. The amplitude of the latter modulates around the ring in a way compensating the amplitude modulations in the next-nearest neighbour tunnelling strengths in the full Hamiltonian. Hence, if left behind can also be detrimental for the rotational symmetry around the ring.}. 
In other words, the broken degeneracy between the angular momentum partners reflects the weak TRS breaking. We observe that the gap within a pair is around $~0.005J-0.007J$ and increases with angular momentum around the ring. 
We note that while its phase can become significant (cf.~Fig.~\ref{fig:nnntunnellingphasemag}), the magnitude of the next-nearest neighbour tunnelling elements, although non-negligible, is 25 to 100 times smaller than the nearest neighbour terms. The relatively small size of these couplings renders the associated gaps to be small as well, which is in general detrimental for topological signatures. 
It would therefore be desirable to investigate ways for amplifying these complex next-nearest neighbour couplings and the hybridization between the delocalized regions to increase the stability of the topological signatures. 
In general, the driving must be sufficiently strong to give rise to significant tunnelling phases in the lattice, as in Fig.~\ref{fig:loc+Chernbig} where the driving strength is $W/\omega\approx 5.6$. The spatial wavelength of the twisting potential, $\beta^{-1}$, can be also tuned where we target values large enough to allow for connected regions, e.g.~as predicted by the conductive ring in Fig.~\ref{fig:avgJ}. To increase the scale of the rings, $\beta$ can be tuned to smaller values with increasing system sizes to induce stable and connected extended regions.

It is important to highlight the difference of our system with the cases considered in the literature so far. Both in periodic crystalline systems (as in the regular Hofstadter~\cite{Hofstadter1976} or Haldane models per unit cell~\cite{HaldaneOriginal}), and in aperiodic lattices such as quasicrystals, one usually considers a uniform magnetic field across the plane which in a sense breaks the TRS in a spatially homogeneous way. While we note that the spatial aperiodicity in the case of a quasicrystal naturally affects the induced flux pattern per each tile, this follows the underlying structure with long-range correlations. 
In contrast in our case, TRS is also broken inhomogeneously on top of a spatially inhomogeneous lattice, which is closer to disordered systems. To this end we now consider a disordered Chern insulator to investigate the effect of this inhomogeneity further.

\begin{figure}
    \centering
    \includegraphics[width=.8\linewidth]{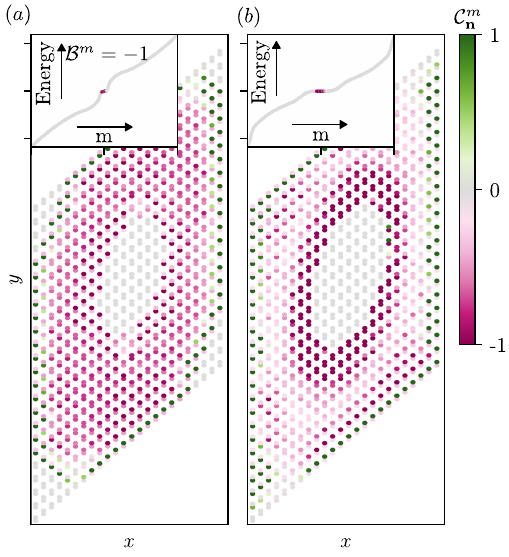}
\caption{The Chern marker of the Haldane model at half filling, confined in an annulus region with radius $R=9$ and width $\Delta R=10$, obtained by setting tunnelling amplitudes to zero outside. (a) Clean system with $\phi= \pi/2$, $J_{nnn}/J_{nn} = 0.01$ and vanishing sublattice offsets. Inset depicts the finite Bott index $B^m =-1$, where the Chern marker also saturates to $-1$ as shown for $m=440$. 
(b) Disordered system where we multiply the hopping amplitudes by $[\Theta(r-R)+\exp(-(r-R)^2/2(\Delta R)^2)]/2$, with the Heaviside step function $\Theta(r)$.  $J_{nnn}$ are chosen randomly uniform between $0$ and $0.01J_{nn}$, and $\phi$ across a wide range between $0$ and $\pi$. Despite the strong disorder, some in-gap states with $B^m=-1$ survive with a ring-shaped Chern marker, here shown for $m=485$. }
    \label{fig:haldane}
\end{figure}
{\it Disordered Haldane model}--- The Haldane model is a paradigmatic example of a system supporting non-trivial Chern numbers under a net zero magnetic field~\cite{HaldaneOriginal}. 
It is defined on a honeycomb lattice with real nearest-neighbour ($J_{nn}$) and complex next-nearest-neighbour ($J_{nnn}e^{\pm i\phi}$) 
tunnelling amplitudes with a phase $\phi$ induced by a non-uniform magnetic field within a unit cell. To draw better contrast with our case, we tune the tunnelling strengths such that the system is confined to an annulus region (of width $\Delta R$) by setting $J_{nn}$ and $J_{nnn}$ to zero outside. 
We consider vanishing sublattice offsets and $\phi=\pi/2$, which at half filling corresponds to a Chern number of -1 for the lower band~\cite{HaldaneOriginal}. In the finite annulus geometry, we obtain Bott index $-1$  for in-gap states as shown in Fig.~\ref{fig:haldane}(a), with the associated Chern marker saturating to $-1$ within the bulk of the lattice as expected.  Note that we here take $J_{nnn}/J_{nn} = 0.01$, at the same order with our system under periodic twisting, which makes the associated gap small, while the Chern phase nonetheless clearly lies within the topologically non-trivial regime~\cite{HaldaneOriginal}.

We would like to study how these signatures evolve when we add disorder to the system. We do this by randomising both the next-nearest-neighbour tunnelling phases and magnitudes by sampling them uniformly across a wide range within the same Chern phase. In particular, we choose $J_{nnn}$ randomly from a uniform distribution between $0$ and $0.01J_{nn}$, and $\phi$ between $0$ and $\pi$. 
The associated Bott and Chern markers are given in Fig.~\ref{fig:haldane}(b). While this disorder diminishes the size of the topological gap significantly, and thus the topological robustness of the states, we observe that some in-gap states still feature a Bott index of $-1$ with the ring-shaped Chern marker surviving. These topological signatures in Fig.~\ref{fig:haldane}(b) are indeed adiabatically connected to the Chern phase of the clean Haldane model in (a), and demonstrate clear resemblance with the properties of the ring states.

\section{Discussion and Conclusions}
We now discuss some experimental realizations. Superimposing optical lattice potentials constitutes a technique widely employed in ultracold atomic systems, for example, to study localization properties of AA models~\cite{Bordia2017} and to create an eightfold-symmetric quasicrystal from two square lattices overlaid at a $45^\circ$ twist angle~\cite{ViebahnSchneider19_PRL_QC}, where the phase and the amplitude of the lattice potentials can be controlled with high precision. The periodic modulation of these lattices, which is mainly implemented by shaking them, has already revealed interesting physical phenomena~\cite{Eckardt2017,BaiWeld2025,DottiWeld25_PRR_1D,PRajagopalWeld19_PRL_phasonic}. As a more direct approach, this tunability of optical lattices can be utilized to modulate the perturbing lattice potentials to induce a periodic twisting effect. Photonic lattices can offer another feasible platform where the twisting can be implemented spatially with periodic modulations along the wave guides~\cite{Bandres16_PRX,Ozawa2019}.

Furthermore, we propose a new way to achieve periodic twisting by taking advantage of quantum gas microscopes, where desired potential landscapes can be realized by using digital micromirror devices with single-site resolution, which can be most importantly reprogrammed on the go~\cite{GrossBakr21_NatPhys}. Namely, instead of continuously twisting a perturbing lattice~\cite{hur2025_arxiv_MBL_DMD}, it can be projected onto the primary lattice holographically in the form of a discrete step-wise drive. For $S$ number of steps within a Floquet period, the twist angle can be kept constant at the value $\varphi_s=\frac{2\pi}{S}s$ during the step $s=1,2,...,S$ for a time $\tau=T/S$ and then changed. The sharp transition between the steps can also be smoothened for experimental implementation, by ramping up/down the potential strengths with a sinusoidal profile. Note that this is in spirit similar to a continuous circular drive that can be accomplished by staggered modulations, which has allowed for realizing novel Floquet topological phases~\cite{quelle17_NJP,Unal19_PRL,wintersperger20_NatPhys}. This discrete holographic twisting is a highly promising avenue and provides more knobs to tune, which can also alter the localization and topological signatures in a way that helps stabilize them further. We leave these investigations for future work.

To summarize, we have shown that the periodic twisting of two square lattices gives rise to rich localization and topological signatures. Periodic twisting introduces a spatially-dependent multi-frequency drive which leads to an inhomogeneous dynamical localization effect that varies with distance from the axis of rotation. Even for on-site potential strengths far exceeding the underlying energy scales of the primary lattice, we have demonstrate the presence of conductive rings hosting extended ring states, which remain delocalized only along quasi-one-dimensional rings. Moreover, we have found that time reversal symmetry is broken locally in the lattice, which allows the Floquet eigenstates to feature non-trivial topological Bott index and Chern marker for these ring states. We have demonstrated that the weak-TRS breaking in combination with hybridization of spatially separated delocalized ring regions gives rise to these topological signatures. We contrast these results with the disordered limit of the Haldane model, where similar topological features can survive despite the addition of drastic disorder to the system. 
Going beyond the static and clean quasicrystal cases under a net uniform magnetic field, we show that periodic twisting can provide a useful knob to study localization and topological signatures in aperiodic systems and shine light onto the origin of unconventional topologically non-trivial states embedded in the bulk. 

Our work opens up multiple avenues for future research. We expect similar physics in one-dimensional chains described by a driven Aubry-André-Harper model, where the driving is induced by periodically changing the spatial frequency of the on-site potential. Periodically modulated frequency would then induce spatially depended driving, which could give rise to localized eigenstates in certain regions of the chain, analogous to our ring localization. A similar model with the phase of the optical potential modulated periodically, instead of its frequency, was implemented in ultracold atoms recently~\cite{Shimasaki2024}, where the localization transition was observed dynamically, which highlights a promising route towards periodic twisting. Investigating the interplay of the peculiar twisting-induced driving we studied here and the presence of many-body interactions would be another interesting avenue for future research, as it could reveal novel collective dynamics and emergent phenomena in aperiodic systems.

\begin{acknowledgments}
	{\it Acknowledgments --} J.W.~thanks Calvin Hooper and Oliver Breach for productive discussions. We also thank Robert-Jan Slager, Nigel Cooper and Mike Gunn for discussions and comments. The work of J.W.~is funded by the European Union (ERC, QuSimCtrl, 101113633).
    The work of A.\v{S}. is supported by the European Union’s Horizon Europe research and innovation programme under the Marie Sk\l{}odowska-Curie Actions Grant agreement No. 101104378. F.N.\"U.~acknowledges support from the Royal Society under Grant No.~URF/R1/241667, the Simons Investigator Award [Grant No.~511029] and Trinity College Cambridge. The work of F.N.\"U.~was performed in part at Aspen Center for Physics, which is supported by National Science Foundation grant PHY-2210452. 
\end{acknowledgments}

\bibliography{UsedReferences}

@PREAMBLE{
 "\providecommand{\noopsort}[1]{}" 
 # "\providecommand{\singleletter}[1]{#1}%" 
}

@article{BM2011,
  doi = {10.1073/pnas.1108174108},
  url = {https://doi.org/10.1073/pnas.1108174108},
  year = {2011},
  month = jul,
  publisher = {Proceedings of the National Academy of Sciences},
  volume = {108},
  number = {30},
  pages = {12233--12237},
  author = {Rafi Bistritzer and Allan H. MacDonald},
  title = {Moir{\'{e}} bands in twisted double-layer graphene},
  journal = {Proceedings of the National Academy of Sciences}
}

@misc{hur2025_arxiv_MBL_DMD,
      author={Junhyeok Hur and Joey Li and Byungjin Lee and Kiryang Kwon and Minseok Kim and Samgyu Hwang and Sumin Kim and Yong Soo Yu and Amos Chan and Thorsten Wahl and Jae-yoon Choi},
      eprint={2508.20699},
      title = {Stability of many-body localization in two dimensions},
      archivePrefix={arXiv},
      url={https://arxiv.org/abs/2508.20699}, 
}

@article{Antao2025_ChernMosaicTN,
  title = {Tensor Network Method for Real-Space Topology in Quasicrystal Chern Mosaics},
  author = {Ant\~ao, Tiago V. C. and Sun, Yitao and Fumega, Adolfo O. and Lado, Jose L.},
  journal = {Phys. Rev. Lett.},
  volume = {136},
  issue = {15},
  pages = {156601},
  numpages = {8},
  year = {2026},
  month = {Apr},
  publisher = {American Physical Society},
  doi = {10.1103/hhdf-xpwg},
  url = {https://link.aps.org/doi/10.1103/hhdf-xpwg}
}

@article{Grover22_NatPhys_ChernMosaic,
  title = {Chern mosaic and Berry-curvature magnetism in magic-angle graphene},
  volume = {18},
  ISSN = {1745-2481},
  url = {http://dx.doi.org/10.1038/s41567-022-01635-7},
  DOI = {10.1038/s41567-022-01635-7},
  number = {8},
  journal = {Nat. Phys.},
  publisher = {Springer Science and Business Media LLC},
  author = {Grover,  Sameer and Bocarsly,  Matan and Uri,  Aviram and Stepanov,  Petr and Di Battista,  Giorgio and Roy,  Indranil and Xiao,  Jiewen and Meltzer,  Alexander Y. and Myasoedov,  Yuri and Pareek,  Keshav and Watanabe,  Kenji and Taniguchi,  Takashi and Yan,  Binghai and Stern,  Ady and Berg,  Erez and Efetov,  Dmitri K. and Zeldov,  Eli},
  year = {2022},
  month = jun,
  pages = {885–892}
}

@article{Lu2019,
  doi = {10.1038/s41586-019-1695-0},
  url = {https://doi.org/10.1038/s41586-019-1695-0},
  year = {2019},
  month = oct,
  publisher = {Springer Science and Business Media {LLC}},
  volume = {574},
  number = {7780},
  pages = {653--657},
  author = {Xiaobo Lu and Petr Stepanov and Wei Yang and Ming Xie and Mohammed Ali Aamir and Ipsita Das and Carles Urgell and Kenji Watanabe and Takashi Taniguchi and Guangyu Zhang and Adrian Bachtold and Allan H. MacDonald and Dmitri K. Efetov},
  title = {Superconductors,  orbital magnets and correlated states in magic-angle bilayer graphene},
  journal = {Nature}
}

@article{Stepanov2020,
  doi = {10.1038/s41586-020-2459-6},
  url = {https://doi.org/10.1038/s41586-020-2459-6},
  year = {2020},
  month = jul,
  publisher = {Springer Science and Business Media {LLC}},
  volume = {583},
  number = {7816},
  pages = {375--378},
  author = {Petr Stepanov and Ipsita Das and Xiaobo Lu and Ali Fahimniya and Kenji Watanabe and Takashi Taniguchi and Frank H. L. Koppens and Johannes Lischner and Leonid Levitov and Dmitri K. Efetov},
  title = {Untying the insulating and superconducting orders in magic-angle graphene},
  journal = {Nature}
}

@article{Song2023,
  author = {Song, Zhenhao and Seifert, Urban F. P. and Luo, Zhu-Xi and Balents, Leon},
  title = {Mott insulators in moiré transition metal dichalcogenides at fractional fillings: Slave-rotor mean-field theory},
  journal = {Phys. Rev. B},
  volume = {108},
  number = {15},
  pages = {155109},
  year = {2023},
  month = oct,
  doi = {10.1103/PhysRevB.108.155109}
}

@article{Deng2025,
  title = {Frozen non-equilibrium dynamics of exciton {M}ott insulators in moiré superlattices},
  author = {Deng, Shibin and Park, Heonjoon and Reimann, Jonas and Peterson, Jonas M. and Blach, Daria D. and Sun, Meng-Jia and Yan, Tengfei and Sun, Dewei and Taniguchi, Takashi and Watanabe, Kenji and Xu, Xiaodong and Kennes, Dante M. and Huang, Libai},
  journal = {Nature Materials},
  year = {2025},
  volume = {24},
  pages = {527-534},
  month = mar,
  doi = {10.1038/s41563-025-02135-8}
}

@article{Song2022,
  title = {Magic-Angle Twisted Bilayer Graphene as a Topological Heavy Fermion Problem},
  author = {Song, Zhi-Da and Bernevig, B. Andrei},
  journal = {Phys. Rev. Lett.},
  volume = {129},
  issue = {4},
  pages = {047601},
  numpages = {10},
  year = {2022},
  month = {Jul},
  publisher = {American Physical Society},
  doi = {10.1103/PhysRevLett.129.047601},
  url = {https://link.aps.org/doi/10.1103/PhysRevLett.129.047601}
}

@article{Yasuda21_Science,
author = {Kenji Yasuda  and Xirui Wang  and Kenji Watanabe  and Takashi Taniguchi  and Pablo Jarillo-Herrero },
title = {Stacking-engineered ferroelectricity in bilayer boron nitride},
journal = {Science},
volume = {372},
number = {6549},
pages = {1458-1462},
year = {2021},
doi = {10.1126/science.abd3230},
URL = {https://www.science.org/doi/abs/10.1126/science.abd3230}}

@article{woods2021_NatComm,
  title={Charge-polarized interfacial superlattices in marginally twisted hexagonal boron nitride},
  author={Woods, CR and Ares, Pablo and Nevison-Andrews, Harriet and Holwill, MJ and Fabregas, Rene and Guinea, Francisco and Geim, AK and Novoselov, KS and Walet, NR and Fumagalli, Laura},
  journal={Nature communications},
  volume={12},
  number={1},
  pages={347},
  year={2021},
  publisher={Nature Publishing Group UK London},
  url={https://www.nature.com/articles/s41467-020-20667-2}
}

@article{Klein2023,
  title = {Electrical switching of a bistable moiré superconductor},
  volume = {18},
  ISSN = {1748-3395},
  url = {http://dx.doi.org/10.1038/s41565-022-01314-x},
  DOI = {10.1038/s41565-022-01314-x},
  number = {4},
  journal = {Nature Nanotechnology},
  publisher = {Springer Science and Business Media LLC},
  author = {Klein,  Dahlia R. and Xia,  Li-Qiao and MacNeill,  David and Watanabe,  Kenji and Taniguchi,  Takashi and Jarillo-Herrero,  Pablo},
  year = {2023},
  month = jan,
  pages = {331–335}
}

@article{Bandres16_PRX,
  title = {Topological Photonic Quasicrystals: Fractal Topological Spectrum and Protected Transport},
  author = {Bandres, Miguel A. and Rechtsman, Mikael C. and Segev, Mordechai},
  journal = {Phys. Rev. X},
  volume = {6},
  issue = {1},
  pages = {011016},
  numpages = {12},
  year = {2016},
  month = {Feb},
  publisher = {American Physical Society},
  doi = {10.1103/PhysRevX.6.011016},
  url = {https://link.aps.org/doi/10.1103/PhysRevX.6.011016}
}

@article{Marra20_PRB_qperThouless,
  title = {Topologically quantized current in quasiperiodic Thouless pumps},
  author = {Marra, Pasquale and Nitta, Muneto},
  journal = {Phys. Rev. Res.},
  volume = {2},
  issue = {4},
  pages = {042035},
  numpages = {5},
  year = {2020},
  month = {Dec},
  publisher = {American Physical Society},
  doi = {10.1103/PhysRevResearch.2.042035},
  url = {https://link.aps.org/doi/10.1103/PhysRevResearch.2.042035}
}

@article{KrausZilberberg12_PRL_1Dqc,
  title = {Topological States and Adiabatic Pumping in Quasicrystals},
  author = {Kraus, Yaacov E. and Lahini, Yoav and Ringel, Zohar and Verbin, Mor and Zilberberg, Oded},
  journal = {Phys. Rev. Lett.},
  volume = {109},
  issue = {10},
  pages = {106402},
  numpages = {5},
  year = {2012},
  month = {Sep},
  publisher = {American Physical Society},
  doi = {10.1103/PhysRevLett.109.106402},
  url = {https://link.aps.org/doi/10.1103/PhysRevLett.109.106402}
}

@article{KrausZilberberg13_PRL_2Dqc,
  title = {Four-Dimensional Quantum {H}all Effect in a Two-Dimensional Quasicrystal},
  author = {Kraus, Yaacov E. and Ringel, Zohar and Zilberberg, Oded},
  journal = {Phys. Rev. Lett.},
  volume = {111},
  issue = {22},
  pages = {226401},
  numpages = {5},
  year = {2013},
  month = {Nov},
  publisher = {American Physical Society},
  doi = {10.1103/PhysRevLett.111.226401},
  url = {https://link.aps.org/doi/10.1103/PhysRevLett.111.226401}
}

@article{Kitagawa10_PRB,
	Author = {Kitagawa, Takuya and Berg, Erez and Rudner, Mark and Demler, Eugene},
	Doi = {10.1103/PhysRevB.82.235114},
	Issue = {23},
	Journal = {Phys. Rev. B},
	Month = {Dec},
	Numpages = {12},
	Pages = {235114},
	Publisher = {American Physical Society},
	Title = {Topological characterization of periodically driven quantum systems},
	Url = {https://link.aps.org/doi/10.1103/PhysRevB.82.235114},
	Volume = {82},
	Year = {2010}}

@article{Rudner13_PRX,
	Author = {Rudner, Mark S. and Lindner, Netanel H. and Berg, Erez and Levin, Michael},
	Doi = {10.1103/PhysRevX.3.031005},
	Issue = {3},
	Journal = {Phys. Rev. X},
	Month = {Jul},
	Numpages = {15},
	Pages = {031005},
	Publisher = {American Physical Society},
	Title = {Anomalous Edge States and the Bulk-Edge Correspondence for Periodically Driven Two-Dimensional Systems},
	Url = {https://link.aps.org/doi/10.1103/PhysRevX.3.031005},
	Volume = {3},
	Year = {2013}}

@article{Rudner20_NatPhysRev,
  title={Band structure engineering and non-equilibrium dynamics in Floquet topological insulators},
  author={Rudner, Mark S and Lindner, Netanel H},
  journal={Nature reviews physics},
  volume={2},
  number={5},
  pages={229--244},
  year={2020},
  publisher={Nature Publishing Group UK London},
  url={https://www.nature.com/articles/s42254-020-0170-z}
}

@article{Eckardt2017,
  author = {Eckardt, Andre},
  title = {Colloquium: Atomic quantum gases in periodically driven optical lattices},
  journal = {Rev. Mod. Phys.},
  volume = {89},
  number = {1},
  pages = {011004},
  year = {2017},
  month = mar,
  doi = {10.1103/RevModPhys.89.011004}
}

@book{Fisher1993,
  author    = {N. I. Fisher},
  title     = {Statistical Analysis of Circular Data},
  year      = {1993},
  publisher = {Cambridge University Press},
  address   = {Cambridge, UK}
}

@article{Strkalj2022,
  author = {Štrkalj, Antonio and Doggen, Elmer V. H. and Castelnovo, Claudio},
  title = {Coexistence of localization and transport in many-body two-dimensional Aubry-André models},
  journal = {Phys. Rev. B},
  volume = {106},
  number = {18},
  pages = {184209},
  year = {2022},
  month = nov,
  doi = {10.1103/PhysRevB.106.184209}
}

@article{AA1980,
  author = {Aubry, Serge and André, Gilles},
  title = {Analyticity breaking and Anderson localization in incommensurate lattices},
  journal = {Ann. Israel Phys. Soc.},
  volume = {3},
  pages = {133--164},
  year = {1980}
}

@article{Cooper2019,
  title = {Topological bands for ultracold atoms},
  author = {Cooper, N. R. and Dalibard, J. and Spielman, I. B.},
  journal = {Rev. Mod. Phys.},
  volume = {91},
  issue = {1},
  pages = {015005},
  numpages = {55},
  year = {2019},
  month = {Mar},
  publisher = {American Physical Society},
  doi = {10.1103/RevModPhys.91.015005},
  url = {https://link.aps.org/doi/10.1103/RevModPhys.91.015005}
}

@article{BiancoResta2011,
  url = {https://doi.org/10.1103/physrevb.84.241106},
  year = {2011},
  month = dec,
  publisher = {American Physical Society ({APS})},
  volume = {84},
  pages = {241106(R)},
  author = {Raffaello Bianco and Raffaele Resta},
  title = {Mapping topological order in coordinate space},
  journal = {Phys. Rev. B}
}

@article{Gottlob25_PRXq,
  title = {Quasiperiodicity Protects Quantized Transport in Disordered Systems Without Gaps},
  author = {Gottlob, Emmanuel and Borgnia, Dan S. and Slager, Robert-Jan and Schneider, Ulrich},
  journal = {PRX Quantum},
  volume = {6},
  issue = {2},
  pages = {020359},
  numpages = {13},
  year = {2025},
  month = {Jun},
  publisher = {American Physical Society},
  url = {https://link.aps.org/doi/10.1103/zvng-w46m}
}

@article{Rosa2021,
  author = {Rosa, Matheus I. N. and Ruzzene, Massimo and Prodan, Emil},
  title = {Topological gaps by twisting},
  journal = {Commun. Phys.},
  volume = {4},
  number = {1},
  pages = {130},
  year = {2021},
  month = jun,
  doi = {10.1038/s42005-021-00630-3}
}

@article{Nixon24_QST,
  title={Individually tunable tunnelling coefficients in optical lattices using local periodic driving},
  author={Nixon, Georgia M and {\"U}nal, F Nur and Schneider, Ulrich},
  journal={Quantum Science and Technology},
  volume={9},
  number={4},
  pages={045030},
  year={2024},
  publisher={IOP Publishing},
  url= {https://iopscience.iop.org/article/10.1088/2058-9565/ad69bb}
}

@article{Cao2018,
  doi = {10.1038/nature26160},
  url = {https://doi.org/10.1038/nature26160},
  year = {2018},
  month = mar,
  publisher = {Springer Science and Business Media {LLC}},
  volume = {556},
  number = {7699},
  pages = {43--50},
  author = {Yuan Cao and Valla Fatemi and Shiang Fang and Kenji Watanabe and Takashi Taniguchi and Efthimios Kaxiras and Pablo Jarillo-Herrero},
  title = {Unconventional superconductivity in magic-angle graphene superlattices},
  journal = {Nature}
}

@article{Cao2018ins,
  title = {Correlated insulator behaviour at half-filling in magic-angle graphene superlattices},
  volume = {556},
  ISSN = {1476-4687},
  url = {http://dx.doi.org/10.1038/nature26154},
  DOI = {10.1038/nature26154},
  number = {7699},
  journal = {Nature},
  publisher = {Springer Science and Business Media LLC},
  author = {Cao,  Yuan and Fatemi,  Valla and Demir,  Ahmet and Fang,  Shiang and Tomarken,  Spencer L. and Luo,  Jason Y. and Sanchez-Yamagishi,  Javier D. and Watanabe,  Kenji and Taniguchi,  Takashi and Kaxiras,  Efthimios and Ashoori,  Ray C. and Jarillo-Herrero,  Pablo},
  year = {2018},
  month = mar,
  pages = {80–84}
}

@article{Johnstone2022,
  author = {Johnstone, Dean and Colbrook, Matthew J. and Nielsen, Anne E. B. and Öhberg, Patrik and Duncan, Callum W.},
  title = {Bulk localized transport states in infinite and finite quasicrystals via magnetic aperiodicity},
  journal = {Phys. Rev. B},
  volume = {106},
  number = {4},
  pages = {045149},
  year = {2022},
  month = jul,
  doi = {10.1103/PhysRevB.106.045149}
}

@article{BallingNielsen_26BLT,
 title = {Identification and properties of topological states in the bulk of quasicrystals},
  volume = {113},
  ISSN = {2469-9969},
  url = {http://dx.doi.org/10.1103/56vs-zwr8},
  number = {7},
  pages = {075134},
  journal = {Phys. Rev. B},
  publisher = {American Physical Society (APS)},
  author = {Balling-Ansø,  Frode and Krogh,  Jeppe Lykke and Lassen,  Ella Elisabeth and Nielsen,  Anne E. B.},
  year = {2026}
}

@article{Hatakeyama89_JPS,
author = {Hatakeyama ,Tetsuo and Kamimura ,Hiroshi},
title = {Fractal Nature of the Electronic Structure of a Penrose Tiling Lattice in a Magnetic Field},
journal = {Journal of the Physical Society of Japan},
volume = {58},
number = {1},
pages = {260-268},
year = {1989},
doi = {10.1143/JPSJ.58.260},

URL = { 
    
        https://doi.org/10.1143/JPSJ.58.260
    
    

}
    
}

@article{Tran15_PRB,
  title = {Topological Hofstadter insulators in a two-dimensional quasicrystal},
  author = {Tran, Duc-Thanh and Dauphin, Alexandre and Goldman, Nathan and Gaspard, Pierre},
  journal = {Phys. Rev. B},
  volume = {91},
  issue = {8},
  pages = {085125},
  numpages = {9},
  year = {2015},
  month = {Feb},
  publisher = {American Physical Society},
  doi = {10.1103/PhysRevB.91.085125},
  url = {https://link.aps.org/doi/10.1103/PhysRevB.91.085125}
}

@article{Fuchs18_PRB,
  title = {Landau levels in quasicrystals},
  author = {Fuchs, Jean-No\"el and Mosseri, R\'emy and Vidal, Julien},
  journal = {Phys. Rev. B},
  volume = {98},
  issue = {16},
  pages = {165427},
  numpages = {13},
  year = {2018},
  month = {Oct},
  publisher = {American Physical Society},
  doi = {10.1103/PhysRevB.98.165427},
  url = {https://link.aps.org/doi/10.1103/PhysRevB.98.165427}
}

@article{Huang18_PRB_QSH,
  title = {Quantum Spin {H}all Effect and Spin Bott Index in a Quasicrystal Lattice},
  author = {Huang, Huaqing and Liu, Feng},
  journal = {Phys. Rev. Lett.},
  volume = {121},
  issue = {12},
  pages = {126401},
  numpages = {7},
  year = {2018},
  month = {Sep},
  publisher = {American Physical Society},
  doi = {10.1103/PhysRevLett.121.126401},
  url = {https://link.aps.org/doi/10.1103/PhysRevLett.121.126401}
}

@article{kitaev2006anyons,
  title={Anyons in an exactly solved model and beyond},
  author={Kitaev, Alexei},
  journal={Annals of Physics},
  volume={321},
  number={1},
  pages={2-111},
  year={2006},
  publisher={Elsevier},
  url = {http://dx.doi.org/10.1016/j.aop.2005.10.005},
  DOI = {10.1016/j.aop.2005.10.005}
}

@article{Edwards1972,
  doi = {10.1088/0022-3719/5/8/007},
  url = {https://doi.org/10.1088/0022-3719/5/8/007},
  year = {1972},
  month = apr,
  publisher = {{IOP} Publishing},
  volume = {5},
  number = {8},
  pages = {807--820},
  author = {J T Edwards and D J Thouless},
  title = {Numerical studies of localization in disordered systems},
  journal = {Journal of Physics C: Solid State Physics}
}

@article{Toniolo2022,
  url = {https://doi.org/10.1007/s11005-022-01602-6},
  year = {2022},
  month = dec,
  publisher = {Springer Science and Business Media {LLC}},
  volume = {112},
  pages = {126},
  author = {Danieale Toniolo},
  title = {On the Bott index of unitary matrices on a finite torus},
  journal = {Lett. Math. Phys.}
}

@article{Hastings2010,
  doi = {10.1063/1.3274817},
  url = {https://doi.org/10.1063/1.3274817},
  year = {2010},
  month = jan,
  publisher = {{AIP} Publishing},
  volume = {51},
  number = {1},
  pages = {015214},
  author = {Matthew B. Hastings and Terry A. Loring},
  title = {Almost commuting matrices,  localized Wannier functions,  and the quantum {H}all effect},
  journal = {Journal of Mathematical Physics}
}

@article{Hastings+Loring2011,
  doi = {10.1016/j.aop.2010.12.013},
  url = {https://doi.org/10.1016/j.aop.2010.12.013},
  year = {2011},
  month = jul,
  publisher = {Elsevier {BV}},
  volume = {326},
  number = {7},
  pages = {1699--1759},
  author = {Matthew B. Hastings and Terry A. Loring},
  title = {Topological insulators and C$\ast$-algebras: Theory and numerical practice},
  journal = {Annals of Physics}
}

@misc{loring2019guide,
      author={Terry A. Loring},
      eprint={1907.11791},
      archivePrefix={arXiv},
      url={https://arxiv.org/abs/1907.11791}, 
}

@article{BlanesMagnus2009,
  doi = {10.1016/j.physrep.2008.11.001},
  url = {https://doi.org/10.1016/j.physrep.2008.11.001},
  year = {2009},
  month = jan,
  publisher = {Elsevier {BV}},
  volume = {470},
  number = {5-6},
  pages = {151--238},
  author = {S. Blanes and F. Casas and J.A. Oteo and J. Ros},
  title = {The Magnus expansion and some of its applications},
  journal = {Physics Reports}
}

@phdthesis{bianco2014chern,
  title={{Chern Invariant and Orbital Magnetization as Local Quantities}},
  author={Bianco, Raffaello},
  year={2014},
  school={Universit{\`a} degli Studi di Trieste},
  url={https://www.openstarts.units.it/server/api/core/bitstreams/dbca3967-2867-440c-b31b-f3616805d99f/content}
}

@book{Stone2009,
  doi = {10.1017/cbo9780511627040},
  url = {https://doi.org/10.1017/cbo9780511627040},
  year = {2009},
  month = jul,
  publisher = {Cambridge University Press},
  author = {Michael Stone and Paul Goldbart},
  title = {Mathematics for Physics}
}

@article{Kolovsky2011Bfield,
  doi = {10.1209/0295-5075/93/20003},
  url = {https://doi.org/10.1209/0295-5075/93/20003},
  year = {2011},
  month = jan,
  publisher = {{IOP} Publishing},
  volume = {93},
  number = {2},
  pages = {20003},
  author = {A. R. Kolovsky},
  title = {Creating artificial magnetic fields for cold atoms by photon-assisted tunneling},
  journal = {EPL}
}

@phdthesis{OpenBCsMunder,
    title    = {Matrix product state calculations for one-dimensional quantum chains and quantum impurity models},
    school   = {Ludwig-Maximilians-Universität München},
    author   = {Münder,Wolfgang},
    year     = {2011}
}

@article{Hofstadter1976,
  title = {Energy levels and wave functions of {B}loch electrons in rational and irrational magnetic fields},
  author = {Hofstadter, Douglas R.},
  journal = {Phys. Rev. B},
  volume = {14},
  issue = {6},
  pages = {2239--2249},
  numpages = {0},
  year = {1976},
  month = {Sep},
  publisher = {American Physical Society},
  doi = {10.1103/PhysRevB.14.2239},
  url = {https://link.aps.org/doi/10.1103/PhysRevB.14.2239}
}

@book{Vanderbilt2018,
  doi = {10.1017/9781316662205},
  url = {https://doi.org/10.1017/9781316662205},
  year = {2018},
  month = oct,
  publisher = {Cambridge University Press},
  author = {David Vanderbilt},
  title = {Berry Phases in Electronic Structure Theory}
}

@article{Bellissard1994,
  title = {The noncommutative geometry of the quantum {H}all effect},
  volume = {35},
  ISSN = {1089-7658},
  url = {http://dx.doi.org/10.1063/1.530758},
  DOI = {10.1063/1.530758},
  number = {10},
  journal = {Journal of Mathematical Physics},
  publisher = {AIP Publishing},
  author = {Bellissard,  J. and van Elst,  A. and Schulz- Baldes,  H.},
  year = {1994},
  month = oct,
  pages = {5373–5451}
}

@article{Dunlap1986,
  doi = {10.1103/physrevb.34.3625},
  url = {https://doi.org/10.1103/physrevb.34.3625},
  year = {1986},
  month = sep,
  publisher = {American Physical Society ({APS})},
  volume = {34},
  number = {6},
  pages = {3625--3633},
  author = {D. H. Dunlap and V. M. Kenkre},
  title = {Dynamic localization of a charged particle moving under the influence of an electric field},
  journal = {Phys. Rev. B}
}

@article{Lignier2007,
  url = {https://doi.org/10.1103/physrevlett.99.220403},
  year = {2007},
  month = nov,
  publisher = {American Physical Society ({APS})},
  volume = {99},
  pages = {220403},
  author = {H. Lignier and C. Sias and D. Ciampini and Y. Singh and A. Zenesini and O. Morsch and E. Arimondo},
  title = {Dynamical Control of Matter-Wave Tunneling in Periodic Potentials},
  journal = {Phys. Rev. Lett.}
}

@article{Titum2016_PRX_AFAI,
  title = {Anomalous Floquet-Anderson Insulator as a Nonadiabatic Quantized Charge Pump},
  author = {Titum, Paraj and Berg, Erez and Rudner, Mark S. and Refael, Gil and Lindner, Netanel H.},
  journal = {Phys. Rev. X},
  volume = {6},
  issue = {2},
  pages = {021013},
  numpages = {20},
  year = {2016},
  month = {May},
  publisher = {American Physical Society},
  doi = {10.1103/PhysRevX.6.021013},
  url = {https://link.aps.org/doi/10.1103/PhysRevX.6.021013}
}

@article{Nathan2019_PRB_AFI,
  title = {Anomalous Floquet insulators},
  author = {Nathan, Frederik and Abanin, Dmitry and Berg, Erez and Lindner, Netanel H. and Rudner, Mark S.},
  journal = {Phys. Rev. B},
  volume = {99},
  issue = {19},
  pages = {195133},
  numpages = {9},
  year = {2019},
  month = {May},
  publisher = {American Physical Society},
  doi = {10.1103/PhysRevB.99.195133},
  url = {https://link.aps.org/doi/10.1103/PhysRevB.99.195133}
}

@article{slager2024_NatCom_ADS,
  title={Non-Abelian Floquet braiding and anomalous Dirac string phase in periodically driven systems},
  author={Slager, Robert-Jan and Bouhon, Adrien and {\"U}nal, F Nur},
  journal={Nature Communications},
  volume={15},
  number={1},
  pages={1144},
  year={2024},
  publisher={Nature Publishing Group UK London},
  url={https://doi.org/10.1038/s41467-024-45302-2}
}

@misc{KozlovLevitov2023_arxiv,
      author={Kirill Kozlov and Grigor Adamyan and Mariia Kryvoruchko and Yelizaveta Kulynych and Leonid Levitov},
      eprint={2311.16967},
      archivePrefix={arXiv},
      url={https://arxiv.org/abs/2311.16967}, 
}

@article{Jagannathan2013,
  title = {An eightfold optical quasicrystal with cold atoms},
  volume = {104},
  ISSN = {1286-4854},
  url = {http://dx.doi.org/10.1209/0295-5075/104/66003},
  number = {6},
  journal = {EPL},
  publisher = {IOP Publishing},
  author = {Jagannathan,  Anuradha and Duneau,  Michel},
  year = {2013},
  month = dec,
  pages = {66003}
}

@article{JiaZheng2022_RBG,
  title = {Effective curved space-time geometric theory of generic-twist-angle graphene with application to a rotating bilayer configuration},
  author = {Ma, Jia-Zheng and Datta, Trinanjan and Yao, Dao-Xin},
  journal = {Phys. Rev. B},
  volume = {105},
  issue = {24},
  pages = {245102},
  numpages = {14},
  year = {2022},
  month = {Jun},
  publisher = {American Physical Society},
  doi = {10.1103/PhysRevB.105.245102},
  url = {https://link.aps.org/doi/10.1103/PhysRevB.105.245102}
}

@article{HuangLiu2019_PRB,
  title = {Moir\'e localization in two-dimensional quasiperiodic systems},
  author = {Huang, Biao and Liu, W. Vincent},
  journal = {Phys. Rev. B},
  volume = {100},
  issue = {14},
  pages = {144202},
  numpages = {12},
  year = {2019},
  month = {Oct},
  publisher = {American Physical Society},
  doi = {10.1103/PhysRevB.100.144202},
  url = {https://link.aps.org/doi/10.1103/PhysRevB.100.144202}
}

@article{ViebahnSchneider19_PRL_QC,
  title = {Matter-Wave Diffraction from a Quasicrystalline Optical Lattice},
  author = {Viebahn, Konrad and Sbroscia, Matteo and Carter, Edward and Yu, Jr-Chiun and Schneider, Ulrich},
  journal = {Phys. Rev. Lett.},
  volume = {122},
  issue = {11},
  pages = {110404},
  numpages = {6},
  year = {2019},
  month = {Mar},
  publisher = {American Physical Society},
  doi = {10.1103/PhysRevLett.122.110404},
  url = {https://link.aps.org/doi/10.1103/PhysRevLett.122.110404}
}

@article{Viebahn2ToneDrive2022,
  title = {Floquet engineering of individual band gaps in an optical lattice using a two-tone drive},
  author = {Sandholzer, Kilian and Walter, Anne-Sophie and Minguzzi, Joaqu\'{\i}n and Zhu, Zijie and Viebahn, Konrad and Esslinger, Tilman},
  journal = {Phys. Rev. Res.},
  volume = {4},
  issue = {1},
  pages = {013056},
  numpages = {16},
  year = {2022},
  month = {Jan},
  publisher = {American Physical Society},
  doi = {10.1103/PhysRevResearch.4.013056},
  url = {https://link.aps.org/doi/10.1103/PhysRevResearch.4.013056}
}

@article{Wang2frequency2023,
  title = {Topological Floquet engineering using two frequencies in two dimensions},
  author = {Wang, Yixiao and Walter, Anne-Sophie and Jotzu, Gregor and Viebahn, Konrad},
  journal = {Phys. Rev. A},
  volume = {107},
  issue = {4},
  pages = {043309},
  numpages = {11},
  year = {2023},
  month = {Apr},
  publisher = {American Physical Society},
  doi = {10.1103/PhysRevA.107.043309},
  url = {https://link.aps.org/doi/10.1103/PhysRevA.107.043309}
}

@article{Szabo2020,
  title = {Mixed spectra and partially extended states in a two-dimensional quasiperiodic model},
  author = {Szab\'o, Attila and Schneider, Ulrich},
  journal = {Phys. Rev. B},
  volume = {101},
  issue = {1},
  pages = {014205},
  numpages = {10},
  year = {2020},
  month = {Jan},
  publisher = {American Physical Society},
  doi = {10.1103/PhysRevB.101.014205},
  url = {https://link.aps.org/doi/10.1103/PhysRevB.101.014205}
}

@article{Eckardt2015,
  doi = {10.1088/1367-2630/17/9/093039},
  url = {https://doi.org/10.1088/1367-2630/17/9/093039},
  year = {2015},
  month = sep,
  publisher = {{IOP} Publishing},
  volume = {17},
  number = {9},
  pages = {093039},
  author = {Andr{\'{e}} Eckardt and Egidijus Anisimovas},
  title = {High-frequency approximation for periodically driven quantum systems from a Floquet-space perspective},
  journal = {New Journal of Physics}
}

@article{HaldaneOriginal,
  title = {Model for a Quantum {H}all Effect without Landau Levels: Condensed-Matter Realization of the "Parity Anomaly"},
  author = {Haldane, F. D. M.},
  journal = {Phys. Rev. Lett.},
  volume = {61},
  pages = {2015},
  numpages = {0},
  year = {1988},
  month = {Oct},
  publisher = {American Physical Society},
  doi = {10.1103/PhysRevLett.61.2015},
  url = {https://link.aps.org/doi/10.1103/PhysRevLett.61.2015}
}

@article{Fracdim,
  title = {Fractal dimensions of wave functions and local spectral measures on the Fibonacci chain},
  volume = {93},
  ISSN = {2469-9969},
  url = {http://dx.doi.org/10.1103/PhysRevB.93.205153},
  number = {20},
  pages = {205153},
  journal = {Phys. Rev. B},
  publisher = {American Physical Society (APS)},
  author = {Macé,  Nicolas and Jagannathan,  Anuradha and Piéchon,  Frédéric},
  year = {2016},
}

@article{Evers2008,
  title = {Anderson transitions},
  author = {Evers, Ferdinand and Mirlin, Alexander D.},
  journal = {Rev. Mod. Phys.},
  volume = {80},
  issue = {4},
  pages = {1355--1417},
  numpages = {0},
  year = {2008},
  month = {Oct},
  publisher = {American Physical Society},
  doi = {10.1103/RevModPhys.80.1355},
  url = {https://link.aps.org/doi/10.1103/RevModPhys.80.1355}
}

@article{Vogl2020,
  title = {Effective Floquet Hamiltonian in the low-frequency regime},
  author = {Vogl, Michael and Rodriguez-Vega, Martin and Fiete, Gregory A.},
  journal = {Phys. Rev. B},
  volume = {101},
  issue = {2},
  pages = {024303},
  numpages = {6},
  year = {2020},
  month = {Jan},
  publisher = {American Physical Society},
  doi = {10.1103/PhysRevB.101.024303},
  url = {https://link.aps.org/doi/10.1103/PhysRevB.101.024303}
}

@article{Rodriguez-Vega2018,
doi = {10.1088/1367-2630/aade37},
url = {https://dx.doi.org/10.1088/1367-2630/aade37},
year = {2018},
month = {sep},
publisher = {IOP Publishing},
volume = {20},
number = {9},
pages = {093022},
author = {Rodriguez-Vega, M and Lentz, M and Seradjeh, B},
title = {Floquet perturbation theory: formalism and application to low-frequency limit},
journal = {New Journal of Physics}
}

@article{Prodan2010,
  title = {Entanglement Spectrum of a Disordered Topological Chern Insulator},
  volume = {105},
  ISSN = {1079-7114},
  url = {http://dx.doi.org/10.1103/PhysRevLett.105.115501},
  number = {11},
  pages = {115501},
  journal = {Phys. Rev. Lett.},
  publisher = {American Physical Society (APS)},
  author = {Prodan,  Emil and Hughes,  Taylor L. and Bernevig,  B. Andrei},
  year = {2010},
  month = sep 
}

@article{Duncan2024,
  title = {Critical states and anomalous mobility edges in two-dimensional diagonal quasicrystals},
  author = {Duncan, Callum W.},
  journal = {Phys. Rev. B},
  volume = {109},
  issue = {1},
  pages = {014210},
  numpages = {11},
  year = {2024},
  month = {Jan},
  publisher = {American Physical Society},
  doi = {10.1103/PhysRevB.109.014210},
  url = {https://link.aps.org/doi/10.1103/PhysRevB.109.014210}
}

@article{Bordia2017,
  title = {Probing Slow Relaxation and Many-Body Localization in Two-Dimensional Quasiperiodic Systems},
  author = {Bordia, Pranjal and L\"uschen, Henrik and Scherg, Sebastian and Gopalakrishnan, Sarang and Knap, Michael and Schneider, Ulrich and Bloch, Immanuel},
  journal = {Phys. Rev. X},
  volume = {7},
  issue = {4},
  pages = {041047},
  numpages = {8},
  year = {2017},
  month = {Nov},
  publisher = {American Physical Society},
  doi = {10.1103/PhysRevX.7.041047},
  url = {https://link.aps.org/doi/10.1103/PhysRevX.7.041047}
}

@article{Sbroscia2020,
  title = {Observing Localization in a 2D Quasicrystalline Optical Lattice},
  author = {Sbroscia, Matteo and Viebahn, Konrad and Carter, Edward and Yu, Jr-Chiun and Gaunt, Alexander and Schneider, Ulrich},
  journal = {Phys. Rev. Lett.},
  volume = {125},
  issue = {20},
  pages = {200604},
  numpages = {5},
  year = {2020},
  month = {Nov},
  publisher = {American Physical Society},
  doi = {10.1103/PhysRevLett.125.200604},
  url = {https://link.aps.org/doi/10.1103/PhysRevLett.125.200604}
}

@article{Gottlob2023,
  title = {Hubbard models for quasicrystalline potentials},
  author = {Gottlob, E. and Schneider, U.},
  journal = {Phys. Rev. B},
  volume = {107},
  issue = {14},
  pages = {144202},
  numpages = {16},
  year = {2023},
  month = {Apr},
  publisher = {American Physical Society},
  doi = {10.1103/PhysRevB.107.144202},
  url = {https://link.aps.org/doi/10.1103/PhysRevB.107.144202}
}

@article{Martinez23_wavepacket,
  title = {Wave-packet dynamics and edge transport in anomalous Floquet topological phases},
  author = {Mart\'{\i}nez, Miguel F. and \"Unal, F. Nur},
  journal = {Phys. Rev. A},
  volume = {108},
  issue = {6},
  pages = {063314},
  numpages = {9},
  year = {2023},
  month = {Dec},
  publisher = {American Physical Society},
  doi = {10.1103/PhysRevA.108.063314},
  url = {https://link.aps.org/doi/10.1103/PhysRevA.108.063314}
}

@article{Shimasaki2024,
  title = {Reversible Phasonic Control of a Quantum Phase Transition in a Quasicrystal},
  author = {Shimasaki, Toshihiko and Bai, Yifei and Kondakci, H. Esat and Dotti, Peter and Pagett, Jared E. and Dardia, Anna R. and Prichard, Max and Eckardt, Andr\'e and Weld, David M.},
  journal = {Phys. Rev. Lett.},
  volume = {133},
  issue = {8},
  pages = {083405},
  numpages = {6},
  year = {2024},
  month = {Aug},
  publisher = {American Physical Society},
  doi = {10.1103/PhysRevLett.133.083405},
  url = {https://link.aps.org/doi/10.1103/PhysRevLett.133.083405}
}

@article{Ozawa2019,
  title = {Topological photonics},
  author = {Ozawa, Tomoki and Price, Hannah M. and Amo, Alberto and Goldman, Nathan and Hafezi, Mohammad and Lu, Ling and Rechtsman, Mikael C. and Schuster, David and Simon, Jonathan and Zilberberg, Oded and Carusotto, Iacopo},
  journal = {Rev. Mod. Phys.},
  volume = {91},
  issue = {1},
  pages = {015006},
  numpages = {76},
  year = {2019},
  month = {Mar},
  publisher = {American Physical Society},
  doi = {10.1103/RevModPhys.91.015006},
  url = {https://link.aps.org/doi/10.1103/RevModPhys.91.015006}
}

@article{DottiWeld25_PRR_1D,
  title = {Measuring a localization phase diagram controlled by the interplay of disorder and driving},
  author = {Dotti, Peter and Bai, Yifei and Shimasaki, Toshihiko and Dardia, Anna R. and Weld, David M.},
  journal = {Phys. Rev. Res.},
  volume = {7},
  issue = {2},
  pages = {L022026},
  numpages = {6},
  year = {2025},
  month = {Apr},
  publisher = {American Physical Society},
  doi = {10.1103/PhysRevResearch.7.L022026},
  url = {https://link.aps.org/doi/10.1103/PhysRevResearch.7.L022026}
}

@article{PRajagopalWeld19_PRL_phasonic,
  title = {Phasonic Spectroscopy of a Quantum Gas in a Quasicrystalline Lattice},
  author = {Rajagopal, Shankari V. and Shimasaki, Toshihiko and Dotti, Peter and Ra\ifmmode \check{c}\else \v{c}\fi{}i\ifmmode \bar{u}\else \={u}\fi{}nas, Mantas and Senaratne, Ruwan and Anisimovas, Egidijus and Eckardt, Andr\'e and Weld, David M.},
  journal = {Phys. Rev. Lett.},
  volume = {123},
  issue = {22},
  pages = {223201},
  numpages = {6},
  year = {2019},
  month = {Nov},
  publisher = {American Physical Society},
  doi = {10.1103/PhysRevLett.123.223201},
  url = {https://link.aps.org/doi/10.1103/PhysRevLett.123.223201}
}

@article{BaiWeld2025,
  author = {Bai, Yifei and Weld, David M.},
  title = {Tunably polarized driving light controls the phase diagram of one-dimensional quasicrystals and two-dimensional quantum {H}all matter},
  journal = {Phys. Rev. B},
  volume = {111},
  number = {11},
  pages = {115163},
  year = {2025},
  month = mar,
  doi = {10.1103/PhysRevB.111.115163}
}

@article{quelle17_NJP,
  title={Driving protocol for a Floquet topological phase without static counterpart},
  author={Quelle, A and Weitenberg, C and Sengstock, K and Smith, C Morais},
  journal={New Journal of Physics},
  volume={19},
  number={11},
  pages={113010},
  year={2017},
  publisher={IOP Publishing},
  url={https://arxiv.org/abs/2406.01445}
}

@article{wintersperger20_NatPhys,
  title={Realization of an anomalous Floquet topological system with ultracold atoms},
  author={Wintersperger, Karen and Braun, Christoph and {\"U}nal, F Nur and Eckardt, Andr{\'e} and Liberto, Marco Di and Goldman, Nathan and Bloch, Immanuel and Aidelsburger, Monika},
  journal={Nat. Phys.},
  volume={16},
  number={10},
  pages={1058--1063},
  year={2020},
  publisher={Nature Publishing Group UK London},
  url={https://www.nature.com/articles/s41567-020-0949-y}
}

@article{Unal19_PRL,
  title = {How to Directly Measure Floquet Topological Invariants in Optical Lattices},
  author = {\"Unal, F. Nur and Seradjeh, Babak and Eckardt, Andr\'e},
  journal = {Phys. Rev. Lett.},
  volume = {122},
  issue = {25},
  pages = {253601},
  numpages = {6},
  year = {2019},
  month = {Jun},
  publisher = {American Physical Society},
  doi = {10.1103/PhysRevLett.122.253601},
  url = {https://link.aps.org/doi/10.1103/PhysRevLett.122.253601}
}

@article{GrossBakr21_NatPhys,
  title={Quantum gas microscopy for single atom and spin detection},
  author={Gross, Christian and Bakr, Waseem S},
  journal={Nat. Phys.},
  volume={17},
  number={12},
  pages={1316--1323},
  year={2021},
  publisher={Nature Publishing Group UK London},
  doi={10.1038/s41567-021-01370-5}
}

@article{Jitomirskaya1999,
 ISSN = {0003486X},
 URL = {http://www.jstor.org/stable/121066},
 author = {Svetlana Ya. Jitomirskaya},
 journal = {Annals of Mathematics},
 number = {3},
 pages = {1159--1175},
 publisher = {Annals of Mathematics},
 title = {Metal-Insulator Transition for the Almost {M}athieu Operator},
 volume = {150},
 year = {1999}
}

@article{Ostlund1983,
author = {Ostlund, Stellan and Pandit, Rahul and Rand, David and Schellnhuber, Hans Joachim and Siggia, Eric D},
doi = {10.1103/PhysRevLett.50.1873},
journal = {Phys. Rev. Lett.},
month = {jun},
number = {23},
pages = {1873--1876},
publisher = {American Physical Society},
title = {{One-Dimensional Schr{\"{o}}dinger Equation with an Almost Periodic Potential}},
volume = {50},
year = {1983}
}

@article{Kohmoto1983,
author = {Kohmoto, Mahito and Kadanoff, Leo P and Tang, Chao},
doi = {10.1103/PhysRevLett.50.1870},
journal = {Phys. Rev. Lett.},
month = {jun},
number = {23},
pages = {1870--1872},
publisher = {American Physical Society},
title = {{Localization Problem in One Dimension: Mapping and Escape}},
volume = {50},
year = {1983}
}

@article{Kohmoto1984,
title = {Cantor spectrum for an almost periodic {S}chrödinger equation and a dynamical map},
journal = {Physics Letters A},
volume = {102},
number = {4},
pages = {145-148},
year = {1984},
issn = {0375-9601},
doi = {https://doi.org/10.1016/0375-9601(84)90928-9},
url = {https://www.sciencedirect.com/science/article/pii/0375960184909289},
author = {M. Kohmoto and Y. Oono}
}

@article{KohmotoSutherland86_PRB,
  title = {Electronic States on a Penrose Lattice},
  author = {Kohmoto, Mahito and Sutherland, Bill},
  journal = {Phys. Rev. Lett.},
  volume = {56},
  issue = {25},
  pages = {2740--2743},
  numpages = {0},
  year = {1986},
  month = {Jun},
  publisher = {American Physical Society},
  doi = {10.1103/PhysRevLett.56.2740},
  url = {https://link.aps.org/doi/10.1103/PhysRevLett.56.2740}
}

@article{Oktel21_PRB,
  title = {Strictly localized states in the octagonal Ammann-Beenker quasicrystal},
  author = {Oktel, M. \"O.},
  journal = {Phys. Rev. B},
  volume = {104},
  issue = {1},
  pages = {014204},
  numpages = {18},
  year = {2021},
  month = {Jul},
  publisher = {American Physical Society},
  doi = {10.1103/PhysRevB.104.014204},
  url = {https://link.aps.org/doi/10.1103/PhysRevB.104.014204}
}

@article{Jagannathan2021,
  title = {The {F}ibonacci quasicrystal: Case study of hidden dimensions and multifractality},
  author = {Jagannathan, Anuradha},
  journal = {Rev. Mod. Phys.},
  volume = {93},
  issue = {4},
  pages = {045001},
  numpages = {37},
  year = {2021},
  month = {Nov},
  publisher = {American Physical Society},
  url = {https://link.aps.org/doi/10.1103/RevModPhys.93.045001}
}

@article{Strkalj2020,
  author    = {V. Goblot and A. {\v{S}}trkalj and N. Pernet and J. L. Lado and C. Dorow and A. Lema{\^{\i}}tre and L. Le Gratiet and A. Harouri and I. Sagnes and S. Ravets and A. Amo and J. Bloch and O. Zilberberg},
  journal   = {Nat. Phys.},
  title     = {Emergence of criticality through a cascade of delocalization transitions in quasiperiodic chains},
  year      = {2020},
  month     = {jun},
  number    = {8},
  pages     = {832--836},
  volume    = {16},
  doi       = {10.1038/s41567-020-0908-7},
  publisher = {Springer Science and Business Media {LLC}},
}

\clearpage
\newpage
\appendix

\section{Gauge Transformation}
\label{section:gauge}
Starting with a generic Hamiltonian $H(t)$, we make a unitary transformation $U(t)$ for a change of basis,
\begin{equation}
    \begin{aligned}
        \ket{\psi} &\rightarrow \ket{\psi'} = \hat{U}(t)^\dagger \ket{\psi}, \\
        \hat{H}(t) &\rightarrow \hat{H}'(t) = \hat{U}^\dagger \hat{H}(t) \hat{U} - i \hat{U}^\dagger \partial_t \hat{U}.
    \end{aligned}
    \label{eqn:gaugetransformrule}
\end{equation}
For the Hamiltonian in Eq.~\eqref{eqn:Hamiltonian}, 
making a local Gauge transformation of the operators $c_n \rightarrow U_n(t) c_n$ where 
\begin{equation}
    U_n(t) = \exp \bigg( -i \int^t V_\mathbf{n}(t')  \text{d}t' c^\dagger_n c_n \bigg) \, ,
\end{equation}
the Galilean term ($U^\dagger \partial_t U$) in Eq.~\eqref{eqn:gaugetransformrule} cancels the on-site potential term in $H(t)$. This leads to
\begin{equation}
    H'(t)= -\sum_{\langle \mathbf{n},\mathbf{n}' \rangle} J_{\mathbf{n} \mathbf{n}'}(t) c^\dagger_\mathbf{n} c_{\mathbf{n}'} \, ,
\end{equation}
in the lattice frame where the effective time-dependent hopping terms are expressed as the difference in the on-site potentials: 
\begin{equation}
    J_{\mathbf{n} \mathbf{n}'}(t) = J\exp \bigg( -i \int_0^t V_{\mathbf{n}}(t')- V_{\mathbf{n}'}(t') \text{d}t' \bigg) \, .
\end{equation}
Performing a Fourier expansion, $V_\mathbf{n}(t) = W\sum s_{\nu, \mathbf{n}} \exp(i\nu\omega t)$, leads to the expression in Eq.~\eqref{eqn:Jefft}, up to a constant phase term that can be gauged out.

\section{Spectral Content}
\label{section:Spectrum}

The Fourier expansion for the time-dependent potential $V(t)$ in Eq.~\eqref{eqn:potential} can be expressed in terms of Bessel functions of the first kind, $\mathcal{J}_n(x)$ \cite{Stone2009}. The following expansions give the spectral content of the driving at a point given by polar coordinates $(R,\theta)$ on the lattice for the different $\Theta$ values considered here with different symmetries. For convenience, we define $2\pi R \beta = \gamma$.

For $\Theta=0$ case, we obtain the following
\begin{align}
\frac{V(\Theta=0,t)}{W} 
&=\cos(\gamma \cos(\theta-\omega t))+\cos \left(\gamma \sin(\theta-\omega t)\right) \nonumber, \\
&=\sum_{\nu=-\infty}^{\infty} 2 \mathcal{J}_{4\nu}(\gamma)\exp(-i4\nu\theta) \exp(i4\nu\omega t),
\end{align}
where in the first line we used the angle addition formulae.
As expected, we recover the $C_4$ symmetry which requires the Floquet frequency $\Omega=4\omega$. 

For $\Theta=\pi/2$, we find that
\begin{align}   \label{eq_app:driving_potential}
\frac{V(\Theta=\pi/2,t)}{W} &= \sin(\gamma \sin(\theta-\omega t))+\sin(\gamma \cos(\theta-\omega t)) \nonumber, \\
&=\sum_{\nu=-\infty}^{\infty} s_\nu \exp(i\nu\omega t)
\end{align}
where 
\begin{equation}
    s_\nu = 
        \begin{cases}
        -i\mathcal{J}_{\nu}(\gamma)(1+i^{\nu})\exp(-i\nu\theta), & \text{if } \nu \text{ is odd},\\
        0, & \text{otherwise}.
    \end{cases}
    \label{eqn:Fourierforsine}
\end{equation}

\begin{figure}[t]
    \centering
    \includegraphics[scale=1]{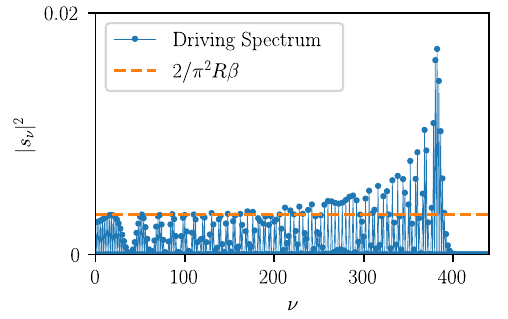}
    \caption{Plot of the Fourier coefficients of $V(t) = \sum_\nu s_\nu \exp(i \nu\omega t)$ for $R=100$ sites away from the origin. High spectral weight is associated with larger frequencies further from the axis of rotation. The orange dashed line denotes the envelope for $\nu \ll 2\pi R \beta$.}
    \label{fig:sinspeccont}
\end{figure}

In Fig.~\ref{fig:sinspeccont} we show the squared Fourier coefficients for the driving on a site at a distance of $R=100$ from the axis of rotation where the multi-frequency nature of the drive is clearly visible. There is a broad range of frequencies with non-negligible power, and a peak in the spectrum occurring at higher frequencies. The spectra are similar for both $\Theta=0$ and $\Theta=\pi/2$ cases, so we focus on the latter in the analysis of the the spectral coefficients. 

When varying $\nu$, we can approximate the Bessel function coefficients with the following asymptotic expansions, that determine the low- and high-frequency regimes,
\begin{equation}
    |s_\nu|^2 \sim 
        \begin{cases}
        \frac{4}{\pi \gamma} \cos\left( \gamma- \frac{\nu\pi}{2}- \frac{\pi}{4}\right), & |\nu^2-1| \ll \gamma \\
         \frac{2}{(\nu!)^2} \bigg( \frac{\gamma}{2} \bigg)^{2\nu}, & \sqrt{\nu+1} \gg \gamma \, .
    \end{cases}
\end{equation}

The transition between the low and high-frequency modes is given approximately by $\nu_c \sim \gamma= 2\pi R \beta$, which follows from the Bessel function in Eq.~\eqref{eqn:Fourierforsine}. The low frequency spectral modes $\nu < \nu_c$ can be approximately modelled by a constant envelope of $2/(\pi^2 R \beta)$ since $J_\nu(2\pi \beta R) \sim 1/\sqrt{2\pi \beta R}$ when $\nu \ll \nu_c$. At very high frequency, the spectral weight decays strongly with a factorial leading to a cut-off. These results are in agreement with Fig.~\ref{fig:sinspeccont}. 

Note that the root mean square (RMS) amplitude of the driving in Eq.~\eqref{eq_app:driving_potential} on a single site is approximately independent of the distance from the origin once the spectrum is broad enough, i.e.~for $2\pi \beta R \gg 1$. This is due to the fact that the number of modes before the cut-off $\nu_c \sim R$ and the squared amplitude of each mode $|s_\nu|^2 \sim R^{-1}$ multiply to a quantity which is independent of the distance from the origin. 

However, when calculating the effective hopping strengths in the system, one must consider not only the induced driving strengths on individual sites, but also the integral of the relative modulations, which becomes the relevant factor: $\phi_{\mathbf{n} \mathbf{n}'}(t) = \int_0^t \text{d}t (V_{\mathbf{n}}(t')- V_{\mathbf{n}'}(t')),$  for $J_{\mathbf{n} \mathbf{n}'}(t) = J \exp(-i\phi_{\mathbf{n} \mathbf{n}'}(t))$. 
The integral of the on-site potential term means that higher frequencies are suppressed by a factor of $1/\omega$ and hence the Peierls phase contributed by the driving is smaller further away from the origin (see Appendix.~\ref{sec:spatvarhop}).

\section{Approximating Spatial Variation of Hopping Amplitudes}
\label{sec:spatvarhop}
We further seek to find the form of the coefficients $\alpha_{\nu,\mathbf{n} \mathbf{n}'}$ for the specific form of our driving with $\Theta=\pi/2$.
The first step is to consider the behaviour of the Fourier modes. We first calculate the time-varying phase of the hopping in Eq.~\eqref{eqn:Jefft} which is just the integrated difference in the on-site potential. Between a pair of sites labelled by $(\mathbf{n},\mathbf{n}')$, the integral in the difference in driving for mode $\nu$ is
\begin{align}
    \int_0^t \Delta V_{\mathbf{n} \mathbf{n}',\nu} \text{d}t'&= A\sum_{\nu=-\infty}^{\infty} \frac{(s_{\nu,\mathbf{n}} - s_{\nu,\mathbf{n}'})}{i\nu} (e^{i\nu\omega t} - 1),\\
    &= C + 2A\sum_{\nu=1}^{\infty} \Im \bigg\{ \frac{(s_{\nu,\mathbf{n}} - s_{\nu,\mathbf{n}'})}{\nu} e^{i\nu\omega t} \bigg\},\\
    &= C+ \phi_{\mathbf{n} \mathbf{n}'}(t),
    \label{eqn:phase}
\end{align}
where $C$ is a constant and we used the fact that $\Delta V$ is real, which allow us to relate the Fourier coefficients of negative modes to positive modes. When we take this phase into an exponential, we can easily gauge away the constant term by making a change of basis. Substituting Eq.~\eqref{eqn:Fourierforsine} for the Fourier coefficients, $s_\nu$, we obtain Eqs.~\eqref{eqn:phasetime} and~\eqref{eqn:alphas} in the main text.

Note that crucially there is no single dominant mode, which means that the drive cannot be described by a single frequency across the whole lattice, highlighting the spatially varying multi-frequency nature. However, there are some general comments we can make about $\alpha_{\nu,\mathbf{n} \mathbf{n}'}$ that follow directly from the properties of Bessel functions. 
At a distance R from the centre of the lattice, we find $\alpha_{\nu,\mathbf{n} \mathbf{n}'} \sim R^{-1/2}$. Taking the phases $\Delta_{\nu, \mathbf{n} \mathbf{n}'}$ from Eq.~\eqref{eqn:phasetime} as random and considering only the envelope of the Bessel functions, we find that
\begin{equation}
    \langle \phi_{\mathbf{n} \mathbf{n}'}^2 \rangle \approx A^2 \bigg|2\sqrt{2}\sqrt{\frac{1}{\pi^2 \beta R}} \bigg|^2 \sum^{\nu_c}_{\nu = \text{ odd}} \nu^{-2} \approx \frac{A^2}{\beta R},
\end{equation}
where the sum over odd values is easily found from the Basel problem solution: $\sum_{\nu=1}^\infty \nu^{-2} = \pi^2/6$. Hence, we find that the RMS value of the time-dependent phase associated with the hopping decays as shown in Eq.~\eqref{eq:RMS} in the main text. 
We hence conclude that, for a given Hamiltonian, the strongest driving effects occur near the rotation axis.

\section{Details of the \texorpdfstring{$\Theta=\pi/2$}{Theta=pi/2} Case}
\label{section:staticsol}

In this section, we discuss some details of the static potential and the time-reversal symmetry in the limit of the vanishing spatial frequency. 

\subsection{Spectrum for Static Potential}
\label{sec:Staticsol}
The static Hamiltonian of a $\Theta=\pi/2$ case is given by
\begin{equation}
    \hat{H} = -J \sum_{\langle \mathbf{n} \mathbf{n'} \rangle} c^\dagger_\mathbf{n} c_\mathbf{n'} + \sum_\mathbf{n} V_\mathbf{n} c^\dagger_\mathbf{n} c_\mathbf{n}, 
\end{equation}
where the potential is given by 
\begin{equation}    \label{eq_app:potential_pi/2}
    V(x,y) = W \left( \sin(2\pi \beta x) + \sin(2\pi \beta y) \right) \, ,
\end{equation}
and we take \mbox{$\beta=2/(1+\sqrt{5})$}.

First, we explore the degeneracy of the  zero-energy, $E=0$, states for a static lattice in two different limits, and on an $N \times N$ lattice with open boundary conditions.

In the limit $W/J \rightarrow \infty$, we end up in the position basis. The line $x=-y$ forms a $V=0$ equipotential and there is a degenerate set of $N$ states with $E=0$ along this diagonal. 

In the opposite limit of $W/J \rightarrow 0$, we can use the previously established results for tight-binding models with open boundary conditions~\cite{OpenBCsMunder}. In this case, there exist a good quantum number $\eta_m=\frac{\pi m}{(N+1)}$ with integer $m$ that is bounded by $1 \leq m \leq N$. 
The eigenvalues of a 1D system are then given by $E_{1D} = -2 J \cos(\eta_m)$. 
In 2D, the $x$ and $y$ directions are separable, as one can see from Eq.~\eqref{eq_app:potential_pi/2}, and the eigenvalues are given as a sum of eigenvalues of each degree of freedom separately, i.e.~$E_{2D}(m_1,m_2) = -2J \left[\cos(\eta_{m_1}) + \cos(\eta_{m_2})\right]$.
Similarly as in the previous limit, we find that the $E=0$ state is $N$-fold degenerate.  

It is surprising that a $E=0$ degeneracy survives even beyond the aforementioned limits, i.e.~when $W/J$ is finite. Furthermore, even more puzzling is the fact that it remains robust to time-dependent driving studied in this work as it is evident from Figs.~\ref{fig:DOSplots2} and~\ref{fig:sinspecexamples}. The robust zero-energy degenerate states have been reported in other aperiodic lattices~\cite{KohmotoSutherland86_PRB,Oktel21_PRB}, and their appearance under periodic driving signals underlying richer dynamics. We leave the investigation of these properties for future work. 

Alongside this extensive degeneracy, there is also particle-hole symmetry in the spectrum. We define the two terms in Eq.~\eqref{eqn:Hamiltonian} as $\hat{H}_J$ for the kinetic and $\hat{V}(t)$ for the potential terms. 
We first choose a basis for our 2D states to be cast into a matrix. We take a “snake-like” basis. This is chosen such that the operator
\begin{equation}
    \hat{A} = \sum_n (-1)^n c_n^\dagger c_n
\end{equation} gives a checkerboard tiling of plus and minus signs. Thus, every site is connected to sites of the opposite sign and the lattice is bipartite. $\hat{H}_J$ is odd under the action of this operator, but $\hat{V}(t)$ is even:
\begin{equation}
    \hat{A} \hat{H} \hat{A} = -\hat{H}_J + \hat{V}(t).
\end{equation}

We can also construct an operator $\hat{B}$, which reflects all the positions about the line $y=-x$. The operator $\hat{B}$ acting on the Hamiltonian gives
\begin{equation}
    \hat{B} \hat{H} \hat{B} = \hat{H}_J - \hat{V}(t),
\end{equation}
since $V$ has reflection antisymmetry for $\Theta=\pi/2$ and the hopping term is invariant under reflection. One can alternatively think of $\hat{B}$ as the equivalent of $\hat{A}$ on a reciprocal space lattice. 

These operators satisfy $\hat{A}^2=\hat{B}^2= \mathbb{1}$. For a general eigenstate $\ket{\varphi_n}$ of energy $E_n$,
\begin{align}
    \hat{A}\hat{B} (\hat{H}_J+\hat{V}(t)) \hat{B} \hat{A}\hat{A}\hat{B}\ket{\varphi_n}
    &=E_n \hat{A}\hat{B} \ket{\varphi_n} \\
    - (\hat{H}_J+\hat{V}(t))\hat{A} \hat{B}\ket{\varphi_n} 
    &= E_n \hat{A}\hat{B} \ket{\varphi_n},
\end{align}
we find that there is always a partner eigenstate $\hat{A} \hat{B} \ket{\varphi_n}$ at $-E_n$ thus proving the symmetry of the spectrum. The reflection anti-symmetry of the on-site potential $\hat{V}$ is key for this to hold. While we have proven it here for the instantaneous eigenstates at some time $t$, it is also observed to hold for the Floquet eigenstates. 

\subsection{Time Reversal Symmetry for \texorpdfstring{$\beta \rightarrow 0$}{beta -> 0}}
In the following, we show that the driving has time-reversal symmetry in the limit of $\beta \rightarrow 0$. In this limit, the potential can be written as
\begin{align}
    V(x,y,t) &= W[x (\cos(\omega t)+\sin(\omega t))\\ &+ y (\cos(\omega t)-\sin(\omega t))]\\
    &= W[x \sin(\omega t+ \pi/4)\\ &+ y \cos(\omega t+\pi/4)].
\end{align}
By inspection, we see that there is no spatially varying phase of the driving, and thus the potential is time-reversal symmetric. It corresponds to a 2D generalization of the potential given in Ref.~\cite{Kolovsky2011Bfield} which is purely real.

More rigorously, we find that the complex part of tunnelling amplitudes, which is necessary (but not sufficient) for breaking the time-reversal symmetry, is zero \cite{Eckardt2017} in this limit. The complex part of the hopping terms between sites labelled by $\mathbf{n}$ and $\mathbf{n}'$ is given by
\begin{equation}
    I_{\mathbf{n}'\mathbf{n}} \propto \int_0^T \sin( -\int_0^t (V_{\mathbf{n}'}(\tau) - V_\mathbf{n}(\tau)) \text{d}\tau +\xi_\mathbf{n}- \xi_{\mathbf{n}'}),
\end{equation}
where the constants $\xi_\mathbf{n}$ represent the local gauge freedom. If we can choose values for the $\xi_\mathbf{n}=0$ such that all the $I_{\mathbf{n}'\mathbf{n}}$ are zero, then the system is time-reversal symmetric. Since $\int_0^t (V_\mathbf{n}'(\tau) - V_\mathbf{n}(\tau))\text{d}\tau$ will just be another harmonic with period $T$, we find that it still averages to zero over the whole period for $\xi_\mathbf{n}=0$. Since $\xi_\mathbf{n}=0$ for all $\mathbf{n}$ satisfies this constraint, the system is time-reversal symmetric. 
We confirm this symmetry numerically by plotting the effective magnetic fluxes. As expected for a system with time-reversal symmetry, it vanishes everywhere to within a high precision of $10^{-10}$. While for finite $\beta$, it is challenging to establish the breaking of time-reversal symmetry analytically due to the inhomogeneous multi-frequency drive, we numerically show the presence of finite local magnetic fluxes in the system as demonstrated in the main text.

\section{Finite-size scaling analysis of localization}
\label{sec:scaling}
In support of the results in the main text, we show a range of system sizes here to demonstrate the localization behaviour concretely. We define the thermodynamic limit with respect to a fixed $\beta$ and adding extra sites around the boundary of the system with the center fixed. This is a useful notion, allowing us to track the same ring states across different system sizes, since they are not rescaled. We provide evidence that the ring states are robust under this scaling definition, in which they remain fixed in size and are therefore localized in the thermodynamic limit.

\begin{figure}[h]
    \centering
    \includegraphics[width=\linewidth]{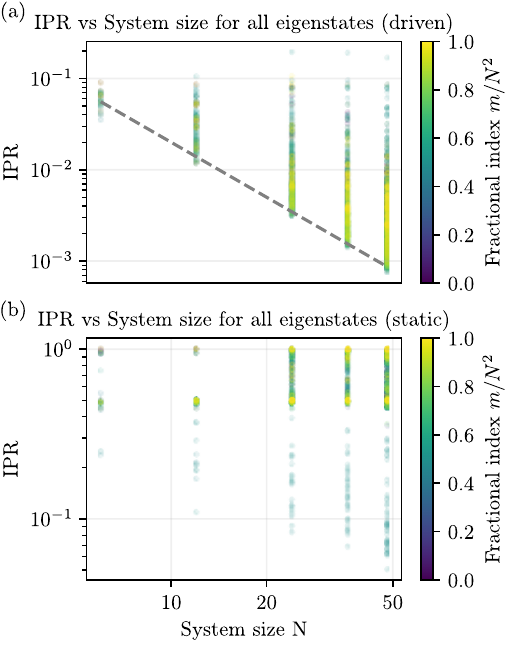}
    \caption{Scatter plots of the IPR across several system sizes for the driven (a) and static (b) case. The dashed line in the driven case shows the trend for a uniform state ($\text{IPR}\sim N^{-2}$); we see the bulk of the states follow this line, indicating delocalization. This is in sharp contrast to a static case in (b) where the bulk of states are localized.}
    \label{fig:IPRscaling}
\end{figure}

First, we give further evidence to the claim of the majority of states being delocalized in the presence of the driving (in contrast to the static case) by showing results from a range of system sizes in Fig.~\ref{fig:IPRscaling}. The driven case in Fig.~\ref{fig:IPRscaling}(a) shows decreasing IPR for the majority of states with system size $N$. This closely follows the dashed line in black of $\text{IPR}\sim 1/N^2$ which corresponds to a delocalized state with weight over extensively many sites. By contrast, in the static case in Fig.~\ref{fig:IPRscaling}(b), we see that at $W=50$, the vast majority of the states are strongly localized. This demonstrates clearly that the drive plays a significant role in delocalizing the states via a dynamical delocalization effect.

Next, we detail the scaling behaviour of the fractal dimension as according to our definition of the thermodynamic limit. First, to identify and study ring states consistently across different system sizes, we identify them as the states that have the maximum value of $\sigma^2_\theta/\sigma^2_R$. In terms of the discrete positions $\mathbf{n}$ for the lattice sites, the radial variance is defined as \begin{equation}
    \sigma^2_R=\sum_\mathbf{n} |\psi_\mathbf{n}|^2|\mathbf{n}|^2-\left( \sum_\mathbf{n} |\psi_\mathbf{n}|^2|\mathbf{n}| \right)^2
\end{equation} with respect to a normalised wavefunction $\psi_\mathbf{n}$. The angular variance requires more care due to the periodicity of $\theta$. It can be defined consistently with respect to the periodicity by defining a vector
\begin{equation}
    \mathbf{S} = \sum_\mathbf{n} |\psi_\mathbf{n}|^2 \, (\cos\theta_\mathbf{n}, \;\sin\theta_\mathbf{n}).
\end{equation}
From this, we can then define the angular variance as
\begin{equation}
    \sigma^2_\theta = 1 - |\mathbf{S}|,
\end{equation}
which has a maximum value of $1$ and minimum value $0$ \cite{Fisher1993}. 

\begin{figure}
    \centering
    \includegraphics[width=1.0\linewidth]{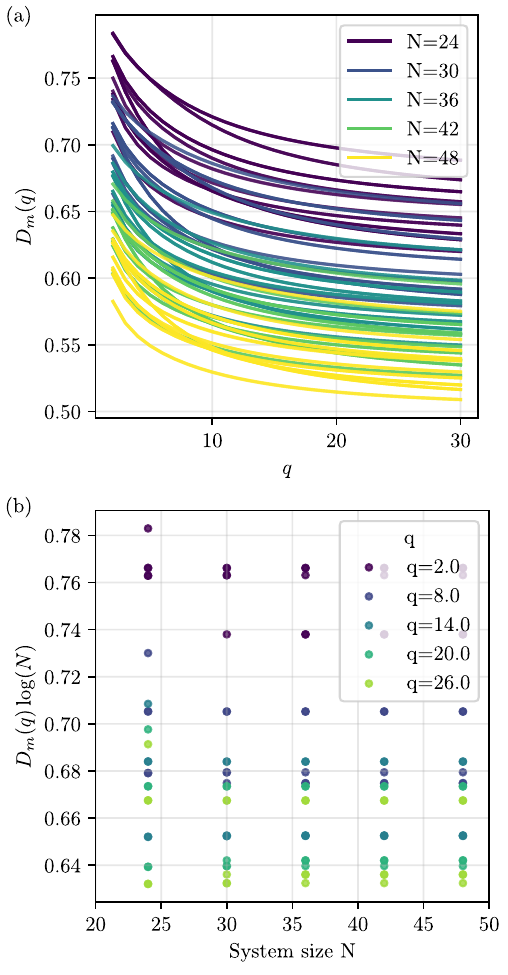}
    \caption{Studying the fractal dimension $D_m(q)$ of ring states with large $\sigma^2_\theta/\sigma^2_R$ across several system sizes. (a) shows the decrease in $D_m(q)$ for 20 different ring states as we increase the size of the system. This is consistent localised behaviour in the thermodynamic limit. (b) shows that the quantity $D_m(q)\log(N)$ is independent of $N$ and leads to a strong scaling collapse (to within fractional error $10^{-5}$). This is directly related to the fact that the rings are very well localized and independent on the system size.}
    \label{fig:DmqL}
\end{figure}

After identifying the ring states across different system sizes, we plot the first few of these in Fig.~\ref{fig:DmqL} to explore the scaling behaviour of the fractal dimension $D_m(q)$. In Fig.~\ref{fig:DmqL}(a), we see that the fractal dimension of the ring state decreases as we go to larger system sizes. We make this result more concrete in Fig.~\ref{fig:DmqL}(b) where we see that multiplying by $\log(N)$ leads to a scaling collapse such that all the $D_m(q)$ look identical, independent of system size, excluding $N=24$ where the close proximity to the boundary affects the behaviour slightly. Hence, we have strong evidence that in the thermodynamic limit of $N \to \infty$, for $m$ corresponding to the ring states, $D_m(q)$ approaches zero logarithmically slowly. 
Physically, $D_m(q)\sim1/\log(N)$ comes from the fact that all the probability density lies within the ring, so when we increase the size of the system the ring effectively appears to shrink to a localized point.

While the ring states are stable to adding more sites within this notion of the thermodynamic limit, the “pocket states”, marked with (\textit{iii}) in Fig.~\ref{fig:ring_states_spectrum}, are not since they rely on the boundary. In Fig.~\ref{fig:ringhybridization}, we show how a rescaling of $\beta$ reduces the radius of the ring states, simultaneously causing a pocket state to turn into an additional ring once its radius becomes smaller than the linear system size $N$.

\section{Topological signatures for \texorpdfstring{$\beta=2/(1+\sqrt{5})=\beta_0$}{beta=2/(1+sqrt(5))=beta0}}

\subsection{Non-trivial Topological Signatures}
\label{sec:Centralisedtop}
\begin{figure}[t!]
    \centering
    \includegraphics[scale=1]{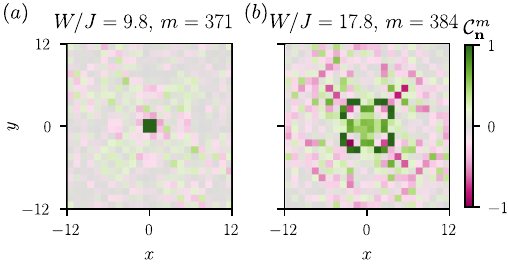}
    \caption{(a) and (b) show examples of Chern marker signatures in the center of the lattice, corresponding to a non-trivial Bott index of $B^m=1$ for the $m$th state specified above the figures.  
    (a) The state features a finite Bott index, however the Chern marker saturates  only at the centre of the lattice over a few sites. Although similar states have been identified as bulk-localized states in eight-fold symmetric quasicrystals under a magnetic field~\cite{Johnstone2022}, the Chern marker signatures remain spatially very restricted.
    (b) Finite Chern marker signatures close to the centre of the lattice.
    These results are for a lattice with $N=24$ with $\omega/J=9$ and $\beta=\beta_0$.}
    \label{fig:Chernmarkerexamples}
\end{figure}

Complementary to the ring signatures explored in the main text, we show some examples of local topological signatures that occupy the center of the lattice in Fig.~\ref{fig:Chernmarkerexamples} (a) and (b). The small extent of these Chern marker signatures is directly tied to the tunnelling coefficients in Fig.~\ref{fig:avgJ} (b) and (c), which is determined by the scale $\beta$.

\begin{figure}[t!]
    \centering
    \includegraphics[width=1\linewidth]{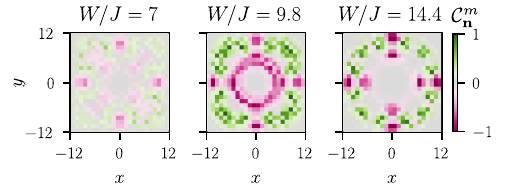}
    \caption{Chern marker and Bott index for the same gap above the $m=561$ state as the disorder potential strength, $W$, is varied. As $W/J$ is tuned, we find that $\mathcal{B}^m=0$ for $W/J=7$ (left), and $\mathcal{B}^m=-1$ for $W/J=9.8$ (middle) and then $\mathcal{B}^m=0$ for $W/J=14.4$ (right). Remaining parameters are the same as Fig.~\ref{fig:Chernmarkerexamples}.}
    \label{fig:Ringb0}
\end{figure}

In Fig.~\ref{fig:Ringb0}, we show how ring-shaped Chern marker signatures can appear and disappear as we tune the system parameters. Non-trivial signatures can persist over a range of values of $W/J$ as can be seen in Fig.~\ref{fig:bottWvary} in the main text. 

\subsection{Ring Transport on \texorpdfstring{$24 \times 24$}{24 X 24} Lattice  \label{sec:ring_transport_24x24}}
\begin{figure}[t!]
    \centering
    \includegraphics[width=1\linewidth]{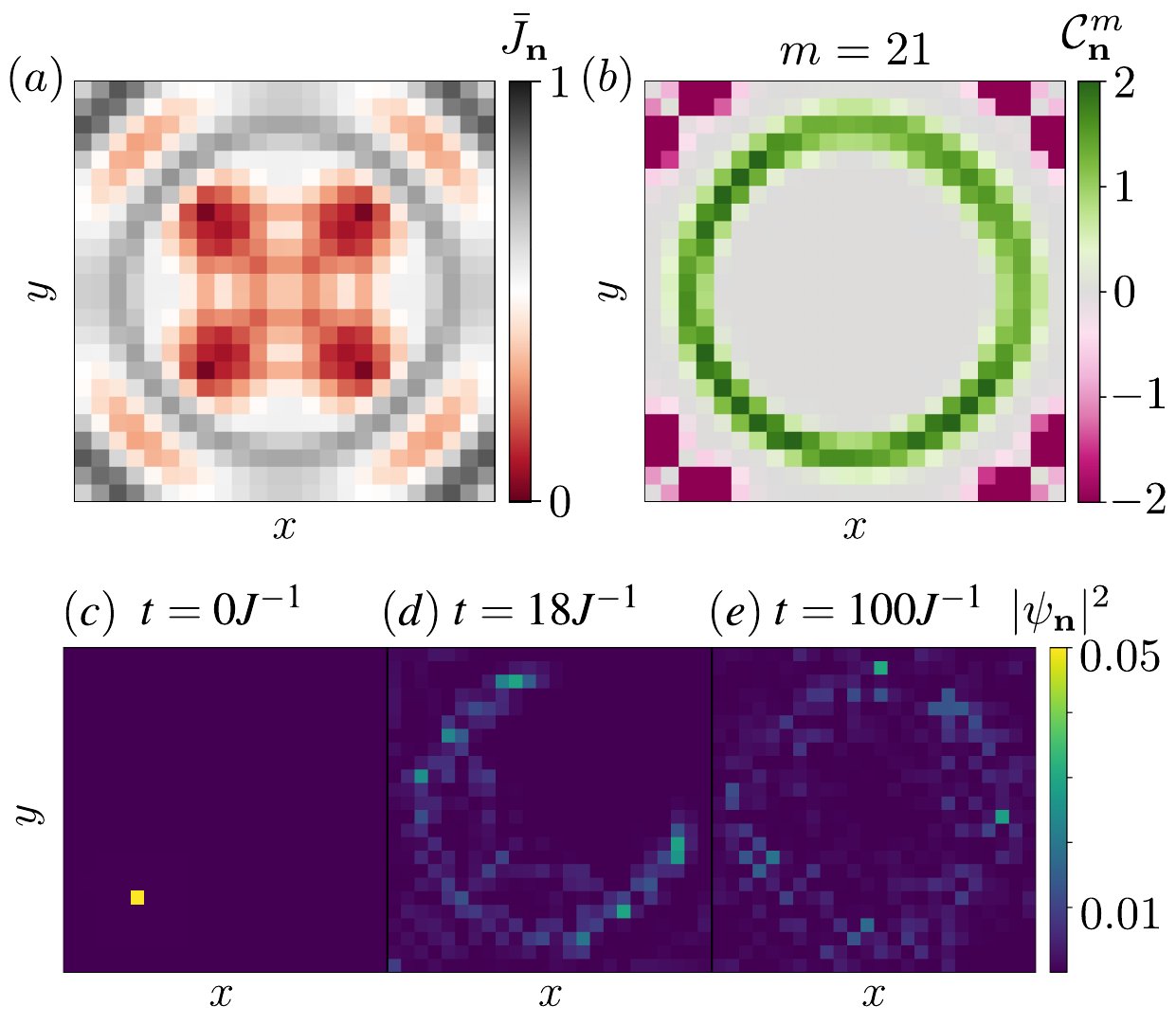}
    \caption{(a) The average tunnelling strength across the lattice as defined in Eq.~\eqref{eq:spatially_averaged_hopping_amplitude}. (b) The Chern marker for $m=21$st state with a Bott index $\mathcal{B}=1$. (c), (d), (e) Snapshots of a wave packet initialized at $t=0$ on the site labelled by $(5,5)$ within the delocalized ring region with large average tunnelling amplitudes in (a). Following the time evolution with the stroboscopic Hamiltonian, a ring-shaped transport signature is visible that persists even at long times. Despite the initial symmetry of the state in (d), we observe a small deviation from the reflection symmetry across the diagonal $x=y$ at long times in (f). The parameters are $W/J=50$, $\omega/J=9$ and $\beta=0.15\beta_0$ for a linear system size of $N=24$.}
    \label{fig:loc+chern}
\end{figure}

Complementary to the topological signatures in Fig.~\ref{fig:loc+Chernbig} shown for a larger system, we can proportionally reduce the scale of the system and recover the same signatures in both Chern marker and transport, even though the localization structures in Fig.~\ref{fig:loc+chern}(a) does not closely resemble Fig.~\ref{fig:avgJ}(c). In contrast to the localized central features demonstrated in the previous subsection, we observe that the ring signatures are more robust to rescaling as can be seen in Fig.~\ref{fig:loc+chern}(b); the behaviour of the lattice is strongly dependent on the finite discretization. In Fig.~\ref{fig:loc+chern}(c), we demonstrate transport dynamics of a wave packet that is initially localised on a single site within the delocalized ring region. While the averaged tunnelling amplitudes in Fig.~\ref{fig:loc+chern}(a) considers only the nearest-neighbour tunnelling processes, the transport dynamics are calculated using the full Floquet Hamiltonian, where the particles show increased transport within the ring.

\section{Flux Gauge Choice}
\label{sec:Appx_fluxgauge}
\begin{figure}
    \centering
    \includegraphics[width=0.99\linewidth]{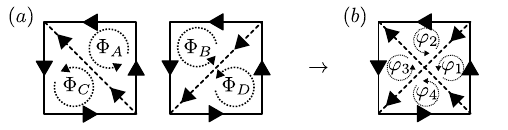}
    \caption{(a) Local fluxes denoted by $\Phi$ within a closed counter-clockwise loop formed by the induced nnn tunnelling amplitudes across the diagonal, which are gauge invariant and hence observable. (b) Flux values within each triangular subdivision of a square plaquette ($\varphi$) are gauge dependent.}
    \label{fig:fluxdef}
\end{figure}

The choice of $\Phi_{\text{triangle}}$ within a square plaquette emerging due to the induced next-nearest neighbour (nnn) tunnelling amplitudes is not unique, similar to the local flux values within a hexagon plaquette in the Haldane model~\cite{HaldaneOriginal}. Namely, there is a family of values that give rise to identical nearest-neighbour (nn) and next-nearest-neighbour (nnn) tunnelling phases, since we can only sample the flux of two triangles at once from a given loop of allowed nn and nnn tunnelling moves on the square lattice as shown in Fig.~\ref{fig:fluxdef}. We denote the flux of the small triangles by $1,2,3,4$, and the flux of the larger triangles by $A, B, C,D$. These fluxes are related by
\begin{align*}
    \Phi_A &= \varphi_1+\varphi_2,\,\Phi_B = \varphi_2+\varphi_3, \\
    \Phi_C &= \varphi_3+\varphi_4,\,\Phi_D = \varphi_4+\varphi_1. 
\end{align*}
If we take $\varphi_{j} \to \varphi_{j} + (-1)^j \delta \varphi$ for arbitrary $\delta \varphi$, the observable fluxes $\Phi_{A,B,C,D}$, are left invariant. This is a local gauge freedom in deciding the fluxes. In the main text, we choose a symmetrized convention where we fix one of the $\varphi$ vanishes, but other conventions are possible and can give rise to different local flux values within triangles. However, we emphasize this does not affect the fact that there are local fluxes present in the system as the fluxes depicted in Fig.~\ref{fig:fluxdef}(a) are gauge invariant. 

\section{Floquet States of Driven Potential with \texorpdfstring{$\Theta=0$}{Theta=0}}
\label{sec:evensymloc}
While we focus on the phase $\Theta=\pi/2$ of the spatial potential in the main text, we now give some details about the $\Theta=0$ case. 
The even symmetry of $\Theta=0$ about the axis of rotation leads to a four-fold rotational symmetry as can be seen in Fig.~\ref{fig:potentialexample}(b). This means that the Floquet frequency, $\Omega$, is four times larger than the underlying rotation frequency, $\Omega=4\omega$. 
In the high-frequency limit of $\Omega \gg \Omega_c$, we can take the first term in the Floquet-Magnus expansion by simply time-averaging the Hamiltonian in Eq.~\eqref{eqn:transformedH} \cite{BlanesMagnus2009}. This approximation is valid to the lowest order in $W/\Omega$, and can be approximately extended to stronger driving by using an extended Hilbert space picture~\cite{Eckardt2015}. 

\begin{figure}[t!]
  \centering
  \includegraphics[scale=1]{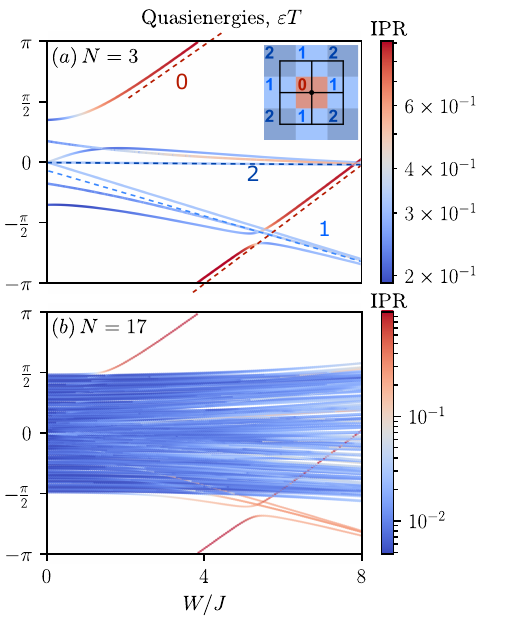}
  \caption{Quasienergy spectrum of Eq.~\eqref{eqn:Hamiltonian} with $\Theta=0$ and $\Omega=4\omega=16J$. 
  (a) The spectrum for a $3\times 3$ lattice, see the inset. Asymptotes (dashed lines) to the average potential values on each set of sites are labelled $W_0, W_1, W_2$ according to their proximity to the origin (see inset for the site labels). 
  (b) The spectrum for a $17\times 17$ lattice. There is a large set of relatively delocalized bands in the bulk where the average potential is weaker for sites further from the origin, while the central sites localize quickly.}
  \label{fig:coslocspec}
\end{figure}

The time average of the on-site potential is 
\begin{equation}
    \langle V_\mathbf{n} \rangle = 2W \mathcal{J}_0(2 \pi \beta R_n),
    \label{eqn:avgpot}
\end{equation}
where $R_n$ is the distance of site $n$ from the axis of rotation and $\mathcal{J}_0(x)$ is the zeroth order Bessel function of the first kind. While there is also a dynamic localization effect in the nearest-neighbour hopping term of the Floquet Hamiltonian, the static on-site potential tends to dominate the localization of states as demonstrated in Fig.~\ref{fig:coslocspec}~\cite{Dunlap1986}.

Since the strongest on-site potential is at and around the center of rotation, from Eq.~\eqref{eqn:avgpot}, the key features of the spectrum can be seen on small lattices as in Fig.~\ref{fig:coslocspec}(a). As the potential strength becomes stronger, the spectrum becomes dominated by the on-site terms alone where the gradients of the asymptotes are given by Eq.~\eqref{eqn:avgpot} with the appropriate $R_n$, i.e.~the behaviour of the spectrum is captured by an effective on-site potential strength $W_n = 2 \mathcal{J}_0(2\pi \beta R_n)$. The flat line of states along the line $W_2\approx -0.019$ is not generic but depends on a particular choice of $\beta=2/(1+\sqrt{5})\equiv\beta_0$. However, the general structure of the spectrum is generic for any $\beta$, just with different asymptotic gradients. Photon-assisted tunnelling between neighbouring sites facilitates hybridization between the states as can be seen at the avoided crossing around $W/J \sim6$ in Fig.~\ref{fig:coslocspec} \cite{Eckardt2017}.

On larger lattices, these key features of the spectrum persist as shown in Fig.~\ref{fig:coslocspec}(b) due to the dominant nature of on-site potentials. The bulk states that appear between $\varepsilon T = \pm \pi/2$ diverge more slowly than the central sites due to the decaying value of the average potential with distance from the origin. The weak strength of the potential further away from the origin results in relatively delocalized states, while the center localizes. 
Ultimately, the physics of the periodic twisting with $\Theta=0$ is dominated by the non-zero average of the potential which leads to localization of the eigenstates across the spectrum in a manner analogous to the static case. This tends to suppress transport across the whole lattice when the driving is sufficiently strong.

\section{Ring hybridization \label{sec:ring_hybridization}}
\begin{figure}[t!]
    \centering
    \includegraphics[width=0.99\linewidth]{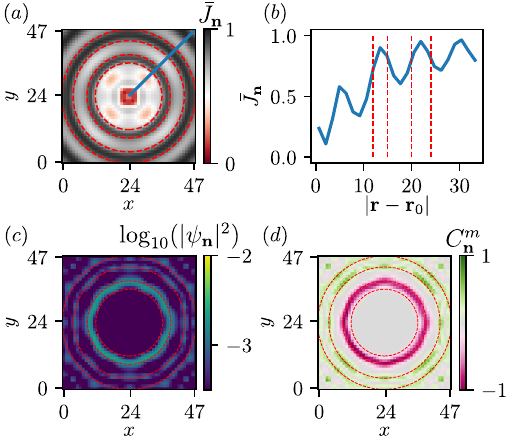}
    \caption{ Periodic twisting with $W=50J,\omega=9J$ in $48\times48$ lattice, where we tune $\beta=0.1\beta_0$ to control the number and position of the delocalized ring regions. (a) The average tunnelling amplitude (Eq.~\eqref{eq:spatially_averaged_hopping_amplitude}), similar to Fig.~\ref{fig:avgJ}, with the values along a diagonal cut (blue line) are given in (b). Large average tunnelling amplitudes corresponding to the conductive rings are highlighted with dashed lines. (c) The probability density of a ring state for $m=199$ with a non-trivial Bott index $\mathcal{B}_m=-1$, where the hybridization between the two ring regions is clearly visible. (d) Corresponding Chern marker signature saturates to -1 within the ring.}
    \label{fig:ringhybridization}
\end{figure}

In Fig.~\ref{fig:RingStHybridz}, we showed how the hybridization between pocket states on the boundary and the rings can result in non-trivial topological signatures. In this appendix, we generalize this to demonstrate hybridization between states of different delocalized regions of the same character by considering rings of different radius being present in the systems.

To achieve this, we modify the value of $\beta=0.1\beta_0$ (i.e.~the ratio of lattice constants of the tight-binding lattice and the perturbing twisted lattice potential) such that multiple conductive rings are induced in the system. We plot the average nearest-neighbour tunnelling amplitudes (Eq.~\eqref{eq:spatially_averaged_hopping_amplitude}) in Fig.~\ref{fig:ringhybridization}(a). We see that $\overline{J}_{\mathbf{n}}$ becomes close to 1 within the second ring (corresponding to the ring in Fig.~\ref{fig:avgJ}), and saturates to unity once more now within a third fully-connected ring region, see dashed lines in Fig.~\ref{fig:ringhybridization}(b) for the average tunnelling amplitudes across a diagonal cut of the lattice. We demonstrate the probability distribution of a ring state in Fig.~\ref{fig:ringhybridization}(c) where the hybridization between two ring regions is clearly visible. Note that due to the difference in the radii of the two rings, their energy scales are different, which can be seen upon approximating with a one-dimensional chain with periodic boundary conditions as discussed in the main text. This implies that only different “angular momentum” modes of the two rings can hybridize. Indeed, the larger ring has a higher energy mode, not only along the polar direction but also along the radial direction. The corresponding Chern marker saturates to the finite Bott index $B^m=-1$ within the ring region, see Fig.~\ref{fig:ringhybridization}(d). This demonstrates the tunability of the platform to control the conductive ring regions.

\onecolumngrid

\setcounter{equation}{0}
\setcounter{figure}{0}
\setcounter{table}{0}
\makeatletter
\renewcommand{\theequation}{S\arabic{equation}}
\renewcommand{\thefigure}{S\arabic{figure}}

\end{document}